\documentclass[12pt]{article}
\usepackage{amssymb,amsmath}
\usepackage{color}
\numberwithin{equation}{section}
\newtheorem{theorem}{Theorem}[section]     
\newtheorem{definition}[theorem]{Definition}
\newtheorem{defi}[theorem]{Definition}
\newtheorem{proposition}[theorem]{Proposition}
\newtheorem{lemma}[theorem]{Lemma}

\newtheorem{example}[theorem]{Example}
\newtheorem{remark}[theorem]{Remark}

\def\d{\partial}
\def\n{\noindent}
\def\f{\frac}
\def\dna{d_{\nabla}}
\def\dlna{d_{L\nabla}}
\def\na{\nabla}
\def\proof{\noindent\hspace{2em}{\itshape Proof: }}
\def\QEDclosed{\mbox{\rule[0pt]{1.3ex}{1.3ex}}} % for a filled box

\def\QED{\QEDclosed} 
\def\endproof{\hspace*{\fill}~\QED\par\endtrivlist\unskip}
\newcommand{\eqa}{\begin{eqnarray}}
\newcommand{\eeqa}{\end{eqnarray}}
\newcommand{\beq}{\begin{equation}}
\newcommand{\eeq}{\end{equation}}

\begin{document}
\title{$F$-manifolds with eventual identities,\\ bidifferential calculus and \\ twisted Lenard-Magri chains}
\author{Alessandro Arsie* and Paolo Lorenzoni**\\
{\small *Department of Mathematics and Statistics}\\
{\small University of Toledo,}
{\small 2801 W. Bancroft St., 43606 Toledo, OH, USA}\\
{\small **Dipartimento di Matematica e Applicazioni}\\
{\small Universit\`a di Milano-Bicocca,}
{\small Via Roberto Cozzi 53, I-20125 Milano, Italy}\\
{\small *alessandro.arsie@utoledo.edu,  **paolo.lorenzoni@unimib.it}}

\date{}
\maketitle
\vspace{-0.2in}
%\begin{center}
%{\em  To Franco Magri  with esteem and friendship in occasion of his 65th birthday}
%\end{center}
{\bf Abstract:} Given an $F$-manifold with eventual identities we examine what this structure entails from the point of view of integrable PDEs
 of hydrodynamic type. In particular, we show that  in the semisimple case the characterization of eventual identities recently
 given by David and Strachan is equivalent to the requirement that $E\,\circ$ has vanishing Nijenhuis torsion. 

Moreover, after having defined new equivalence relations for connections compatible with respect to the $F$-product $\circ$,
 namely {\em hydrodynamically almost equivalent} and {\em hydrodynamically equivalent} connections,
 we show how these two concepts manifest themselves in several specific situations. 

In particular, in the case of an $F$-manifold endowed with eventual identity and two almost hydrodynamically equivalent flat connections
 we are able to derive the recurrence relations for the flows of the associated integrable hierarchy. If the two connections originate from
 a flat pencil of metrics these reduce to the standard bi-Hamiltonian recursion. 

Furthermore, using the geometric set-up proposed here we show how the recurrence relations of the principal hierarchy introduced
 by Dubrovin arise in this general framework and we provide a general cohomological set-up for the conservation laws of the
 semihamiltonian hierarchy associated to a semisimple $F$-manifold with compatible connection and eventual identity. 

Therefore, the point of view we propose, not only highlight the conceptual unity of two well-known recursive schemes 
(principal hierarchy and classical bi-Hamiltonian) but it also provides a far reaching generalization of these recursions that
 relies on the presence of an eventual identity.

\section{Introduction}
In the last twenty years, the interplay between the presence of a Hamiltonian framework and integrability on one hand and geometric structures on the other has been the focus of an intense study. In particular, this has been pursued for a vast class of systems of quasilinear PDEs, usually identified in the literature as systems of hydrodynamic type. Indeed, starting from the first pioneering works of Dubrovin and Novikov it has become more and more apparent that the study of these systems leads naturally to some classical problems in Riemannian geometry. 

More recently,  especially thanks to Dubrovin's works, new light has been shed on this area. Some geometric structures, the so called Frobenius manifolds that were introduced in the study of topological field theory appear naturally within the framework of integrable PDEs of hydrodynamic type. 

Our paper belongs to this research area; however, pursuing a point of view already introduced in  \cite{LPR} and \cite{LP}, we will not follow the usual approach that emphasizes the role played by Riemannian geometry and Hamiltonian structures. Instead, in the approach adopted here, the pivotal role is assigned to a class of symmetric connections, not necessarily originating from a metric, defined on $F$-manifolds. The latter constitute a class of manifolds introduced by Hertling and Manin to generalize Frobenius manifolds. 

To any $F$-manifold it is possible to associate integrable systems of hydrodynamic type. It turns out that the integrability conditions for such systems correspond to the following geometric condition \cite{LPR}:
$$R^k_{lmi}c^n_{pk}+R^k_{lip}c^n_{mk}
+R^k_{lpm}c^n_{ik}=0$$
which expresses a constraint combining the curvature $R^i_{jkl}$ of the connection with the structure constants $c^i_{jk}$ of the associative, commutative product $\circ$ defined on the tangent spaces of $F$-manifolds. 

More specifically, in this work we focus our attention to $F$-manifolds endowed with additional structures or properties, namely the presence of eventual identities or the zero curvature condition and we study how the presence of these structures impacts the corresponding integrable systems of hydrodynamic type. 

It turns out that the most interesting cases correspond to an $F$-manifold endowed with flat connections that are ``hydrodynamically equivalent" or ``almost hydrodynamically equivalent" in a quite peculiar sense, detailed in Section \ref{equivalentsection}.
Indeed, starting from this class of $F$-manifolds it is possible to define recursively the flows of integrable systems of hydrodynamic type.  
In this way, one gets two recursion schemes: the first one, that corresponds to ``hydrodynamically equivalent" connections, is strictly related to the recursion scheme appearing in the principal hierarchy (see Section \ref{recursionprincipal}); the second one, which corresponds to the case of two ``almost hydrodynamically equivalent" connections and the presence of an eventual identity $E$ will provide us with a generalization of the usual Lenard-Magri system, what we call a  twisted Lenard-Magri chain. This includes the classical bi-Hamiltonian recurrence as a special case.
 
Therefore, the point of view we propose, not only highlight the conceptual unity of two well-known recursive schemes (principal hierarchy and classical bi-Hamiltonian) but it also provides a far reaching generalization of these recursions that relies on the presence of an eventual identity. 

For the sake of readability, we detail the organization of the paper, highlighting the results of each section. 
In Section 2 we recall an extension of the usual Fr\"olicher-Nijenhuis bicomplex to differential forms with value in tangent bundle. In the case in which a manifold $M$ is endowed with a flat connection $\nabla$ and with a tangent bundle endomorphism $L$ with vanishing Nijenhuis torsion, this extension will provide a bi-differential complex $d_{\nabla}$, $d_{L\nabla}$ on $\Omega^*(M, TM)$ and in general the operator $d_{\nabla}$ and $d_{L\nabla}$ will be essential to express in an intrinsic way many of the constructions we are going to perform. 

In the brief Section 3 we review Tsarev's theory of semi-Hamiltonian systems, namely the class of integrable systems attached to the geometric structures we are going to explore. 

Section 4 deals with $F$-manifolds with an eventual identity $E$; after reviewing some properties, we show that the recently discovered condition (see \cite{DS}) that characterizes eventual identities among invertible vector fields is equivalent to the endomorphism $V:=E\,\circ$ having zero Nijenhuis torsion in the case the $F$-manifold is semisimple and the eigenvalues of $V$ are distinct. Moreover we prove that the condition that characterizes eventual identity always implies that the corresponding endomorphism $V:=E\, \circ$ has zero Nijenhuis torsion. 

In Section 5 we review the concept of $F$-manifold with compatible connection and in particular we start to explore the interplay of this structure and the presence of an eventual identity $E$. More specifically, we prove that under the condition of semisimplicity of the $F$-manifold and of functional independence of the components of $E$ in the canonical coordinates for $\circ$, the canonical coordinates $\tilde u^i$ for $*$ and the canonical coordinates $u^i$ for $\circ$ are related through a simple reparametrization $\tilde u^i=\tilde u^i(u^i)$. Besides we show that we if write the endomorphism $L:=E\,\circ$ in the canonical coordinates for $*$, it is still diagonal with distinct eigenvalues. 
 All our subsequent investigations are heavily based upon these observations.  

In the short but fundamental Section 6, we introduce the definition of {\em hydrodynamically equivalent} and {\em hydrodynamically almost equivalent} connections on a semisimple $F$-manifold with eventual identity $E$. It turns out that hydrodynamically almost equivalent connections are precisely those that define the same semi-Hamiltonian hierarchy. 

Section 7 deals with conservation laws for the semi-Hamiltonian hierarchy associated to a semisimple $F$-manifold with compatible connection $\nabla$ and eventual identity $E$. We show that the compatibility conditions for the equation defining densities of conservation laws follows from the definition of compatible connection. Furthermore, in the case of an $F$-manifold with compatible flat connection, we identify the recursion relations obeyed by densities of conservation laws and we prove that if the $F$-manifold is endowed with a second flat connection compatible with the multiplicative structure $*$ originating from $E$, then these densities of conservation laws obey an additional system of recursion relations. 

In Section 8 we explore in detail the cohomological nature of the equations determining the symmetries for a semi-Hamiltonian hierarchy; in particular the results obtained will be fundamental in constructing in Section 10, a twisted Lenard-Magri chain associated to a semisimple $F$-manifold with eventual identity $E$ and two almost hydrodynamically equivalent connections $\nabla^{(1)}$ and $\nabla^{(2)}$. We also analyze the special case that arises assuming $\nabla$ to be flat and the corresponding recursion relations that appear to be those of the principal hierarchy. This analysis is completed in Section 9, where the recursion relations of the newly obtained Lenard-Magri chain are compared to those of the principal hierarchy. 

The final Section 10 deals with building a twisted Lenard-Magri chain associated to a semisimple $F$-manifold with eventual identity $E$ and two almost hydrodynamically equivalent connections $\nabla^{(1)}$ and $\nabla^{(2)}$. This chain constitutes a genuine generalization of the classical bi-Hamiltonian recursion relations and it is essentially based on the presence of an eventual identity $E$. We also show that when the two hydrodynamically almost equivalent connections are associated to a flat pencil of metrics, the corresponding twisted chain reduces to the classical bi-Hamiltonian scheme.

\section{An extended Fr\"olicher-Nijenhuis bicomplex}
In this section we recall an extension of the usual Fr\"olicher-Nijenhuis bicomplex to differential forms with value in tangent bundle.

Recall the definition of operators $d$ and $d_L$ on $\Omega^k(M)$:
$$(d \omega)(X_0, \dots, X_k)=\sum_{i=0}^k (-1)^i X_i(\omega(X_0, \dots, \hat{X}_i, \dots, X_k))+$$
$$+\sum_{0\leq i<j\leq k}(-1)^{i+j}\omega([X_i, X_j], X_0, \dots, \hat{X}_i, \dots, \hat{X}_j, \dots X_k),$$
where $X_i(\omega(X_0, \dots, \hat{X}_i, \dots, X_k))$ denotes the action of the vector field $X_i$ on the function $\omega(X_0, \dots, \hat{X}_i, \dots, X_k)$,
 $$(d_L \omega)(X_0, \dots, X_k)=\sum_{i=0}^k (-1)^i (LX_i)(\omega(X_0, \dots, \hat{X}_i, \dots, X_k))+$$
$$+\sum_{0\leq i<j\leq k}(-1)^{i+j}\omega([X_i, X_j]_L, X_0, \dots, \hat{X}_i, \dots, \hat{X}_j, \dots X_k),$$
where $(LX_i)(\omega(X_0, \dots, \hat{X}_i, \dots, X_k))$ indicates the action of the vector field $LX_i$ obtained applying the endomorphism $L$ to $X_i$ and 
$[X_i,X_j]_L=[LX_i, X_j]+[X_i, LX_j]-L[X_i,X_j].$
According to the theory of Fr\"{o}licher-Nijenhuis 
 \cite{FN}, if $L$ is torsionless, namely if 
\beq\label{nijenhis}
[LX,LY]-L\,[X,LY]-L\,[LX,Y]+L^2\,[X,Y]=0 \quad \text{ for any } X, \; Y \; \text{ vector fields},
\eeq
 then
\begin{eqnarray*}
d\cdot d_L+d_L\cdot d=0\quad \text{ and } \quad d_L^2=0.
\end{eqnarray*}
Notice that \eqref{nijenhis}
can be written alternative as
\beq\label{nijenhisalternative}
[LX,LY]=L[X,Y]_L  \quad \text{ for any } X, \; Y \; \text{ vector fields.}
\eeq

We can extend the operators $d$ and $d_L$ to differential $k$ forms with value in $TM$. The extension of $d$, $\dna$ is classical and is known in the literature as {\em exterior covariant derivative} (see for instance \cite{Lee}), while the extension of $d_L$, $\dlna$  is a kind of ``twisted" exterior covariant derivative and it has not appeared before in the literature to the best of our knowledge. Here the introduction of these operators is motivated by the theory of integrability for systems of PDEs of hydrodynamic type. 

We have the following definition for $\dna$:
$$(\dna \omega)(X_0, \dots, X_k)=\sum_{i=0}^k (-1)^i \na_{X_i}(\omega(X_0, \dots, \hat{X}_i, \dots, X_k))+$$
$$+\sum_{0\leq i<j\leq k}(-1)^{i+j}\omega([X_i, X_j], X_0, \dots, \hat{X}_i, \dots, \hat{X}_j, \dots X_k),$$
where $\na_{X_i}$ denotes covariant derivative along $X_i$ of the vector field $(\omega(X_0, \dots, \hat{X}_i, \dots, X_k))$.
The following easy proposition summarizes its main properties:
\begin{proposition}\label{dna}
The following holds:
\begin{enumerate}
\item The operator $\dna$ coincides with $d$ when restricted to scalar valued forms. 
\item If $\omega$ is a $0$-form with value in $TM$, namely a vector field, then $(\dna \omega)(X)=\na_X \omega$, where $X$ is any vector field. 
\item If the connection $\nabla$ is flat, then $\dna\circ \dna=0$ identically. 
\end{enumerate}
\end{proposition}
\proof 
The only point not completely trivial is the third one. To see this is true, simply choose a coordinate system $\{x^1,\dots, x^n\}$ in which $\na_{\d_i}=\d_i$, where $\d_i=\f{\d}{\d x^i}$. This is possible since the connection is flat. Now in this coordinate system, we write the formula for $\dna \omega$ 
$$(\dna \omega)^l_{i_0 \dots i_k}=\sum_{j=0}^k (-1)^j\; \d_{i_j} (\omega^l_{i_0\dots \hat{i}_j \dots i_k})+\sum_{0\leq m<j\leq k}(-1)^{m+j}\; \omega^l_{h \; i_0\dots \hat{i}_m\dots \hat{i}_j \dots i_k}[\d_{i_m}, \d_{i_j}]^h.$$
The second sum vanish identically because $[\d_{i_m}, \d_{i_j}]=0$, while the first sum is just $d\omega^l$, namely $d$ applied to each single component of the vector valued form $\omega$. So the use of flat coordinates decouples the various components and $\dna$ acts on $\omega^l$ like $d$ would act on a collection of $k$-forms. So 
$$(\dna \omega)^l_{i_0 \dots i_k}=(d\omega^l)_{i_0 \dots i_k},$$
in flat coordinates and from this it follows immediately that $\dna\circ \dna=0$ identically. For a coordinate free proof and for more information see \cite{Lee}.
\endproof

We can obtain a new differential $\dlna$ twisting $\dna$ with a $(1,1)$-tensor field, namely an endomorphism of the tangent bundle. We have the following 
\begin{definition}
Given a $(1,1)$-tensor field on a manifold $M$ endowed with a connection $\nabla$, we define the {\em $L$-exterior covariant derivative} $\dlna$ acting on $\Omega^*(M,TM)$ as follows
$$(\dlna \omega)(X_0, \dots, X_k)=\sum_{i=0}^k (-1)^i \na_{LX_i}(\omega(X_0, \dots, \hat{X}_i, \dots, X_k))+$$
$$+\sum_{0\leq i<j\leq k}(-1)^{i+j}\omega([X_i, X_j]_L, X_0, \dots, \hat{X}_i, \dots, \hat{X}_j, \dots X_k).$$
\end{definition}
This definition can obviously be extended to forms with value in sections of a vector bundle $E$ over $M$ as long as we have a connection $\nabla^E$ that enables us to covariantly differentiate sections of $E$ over $M$ (while $L$ is always required to be an endomorphism of $TM$). 

The following proposition summarizes the main properties of $\dlna$.
\begin{proposition}\label{dlna} For the $L$-exterior covariant derivative the following holds: 
\begin{enumerate}
\item $\dlna$ coincides with $d_L$ when restricted to scalar valued differential forms. 
\item If $\omega$ is a $0$-form with value in $TM$, namely a vector field, then $(\dlna \omega)(X)=\na_{LX} \omega$, where $X$ is any vector field. 
\item If $L$ is the identity endomorphism $\dlna$ coincides with $\dna$. 
\item If the connection $\nabla$ is flat and if $L$ has zero Nijenhuis torsion, then $\dlna\circ \dlna=0$ identically. 
\item The operator $\dlna$ is linear in $L$, namely if $M$ and $L$ are two $(1,1)$-tensor fields, then $d_{(L+M)\nabla}=\dlna+d_{M\nabla}.$
\end{enumerate}
\end{proposition}
\proof The first, second and third items are immediate. For the fourth item, we reason as follows. Choose a flat coordinate system $\{x^1, \dots, x^n\}$ where $\nabla_{\d_i}=\d_i$. Then in this coordinate system $\nabla_{L\d_i}=\nabla_{L_i^j \d_j}=L_i^j\nabla_{\d_j}=L_i^j\d_j$. Now let's compute $(\dlna \omega)^l_{i_0\dots i_k}$ in this coordinate system, where $\omega\in \Omega^k(M, TM)$. 
We have 
$$(\dlna \omega)^l_{i_0\dots i_k}=\sum_{j=0}^k(-1)^j L^p_{i_j} \d_p \omega^l_{i_0\dots \hat{i}_j \dots i_k}+\sum_{0\leq m<j\leq k}(-1)^{m+j}\omega^l_{h\; i_0\dots \hat{i}_m\dots \hat{i}_j \dots i_k}[\d_{i_m}, \d_{i_j}]^h_L,$$
where $[\d_{i_m}, \d_{i_j}]^h_L$ denotes the $h$ component of $[\d_{i_m}, \d_{i_j}]_L$. Again, this is just the coordinate expression of $d_L$ acting separately on each form $\omega^l$, the vector components of $\omega$. So we can think that in flat coordinates a vector valued form is just a collection of forms. Since $L$ has zero Nijenhuis torsion, we know that $d_L \circ d_L=0$ identically. Moreover since we proved 
$$(\dlna \omega)^l_{i_0\dots i_k}=(d_L \omega^l)_{i_0\dots i_k},$$
in flat coordinates, then $\dlna\circ\dlna=0$ identically follows  from $d_L\circ d_L=0$. 
The fifth point is immediate due to the linearity of the covariant derivative $\nabla_{(L+M)X}=\nabla_{LX}+\nabla_{MX}$ and the fact that $[X,Y]_{L+M}=[X,Y]_L+[X,Y]_M$. 
\endproof

Using Proposition \ref{dlna} we have the following:
\begin{theorem}\label{bicomplex}
Let $M$ be a manifold endowed with a flat connection $\nabla$ on $TM$ and with two endomorphisms $L$ and $M$ whose Nijienhuis torsion vanishes. Assume that the Nijenhuis torsion of $L+M$ also vanishes. Then $(\Omega^*(M, TM), d_{L\nabla}, d_{M\nabla})$ is a bidifferential complex, namely $d_{L\nabla}\circ d_{L\nabla}=0$, $d_{M\nabla}\circ d_{M\nabla}=0$ and $d_{L\nabla}\circ d_{M\nabla}+d_{M\nabla}\circ d_{L\nabla}=0$.
\end{theorem}
\proof Since $\nabla$ is flat and $Q:=L+M$ has zero Nijenhuis torsion, we have by Proposition \ref{dlna}
$d_{Q\nabla}\circ d_{Q\nabla}=0$. On the other hand, always by Proposition \ref{dlna} $d_{Q\nabla}= d_{L\nabla}+d_{M\nabla}$ and so 
$ 0=(d_{M\nabla}+d_{L\nabla})\circ( d_{M\nabla}+d_{L\nabla})=d_{M\nabla}\circ d_{M\nabla}+d_{L\nabla}\circ d_{M\nabla}+d_{M\nabla}\circ d_{L\nabla}+d_{L\nabla}\circ d_{L\nabla}.$ But since $L$ and $M$ have separately zero Nijenhuis torsion, we see that $d_{M\nabla}\circ d_{M\nabla}=0$, $d_{L\nabla}\circ d_{L\nabla}=0$ and from the previous condition we get the anticommutativity of $d_{L\nabla}$ and $d_{M\nabla}$: $d_{L\nabla}\circ d_{M\nabla}+d_{M\nabla}\circ d_{L\nabla}=0.$
\endproof
In particular, we apply the previous theorem to the case in which $M$ is the identity endomorphism. Indeed one has the following
\begin{lemma}
If $L$ is a torsionless endomorphism of $TM$, then the pencil $Q_{\lambda, \mu}:=\lambda L+\mu I$ is also torsionless for all values of $\lambda$ and $\mu$, where $I$ is the identity endomorphism.  
\end{lemma}
\proof
By \eqref{nijenhisalternative}, we have to prove that $[Q_{\lambda, \mu}X, Q_{\lambda, \mu} Y]=Q_{\lambda, \mu}[X,Y]_{Q_{\lambda, \mu}}$ for all vector fields $X, Y$. Expanding this expression we find
$$\lambda^2[ LX, L Y]+\mu \lambda [ X,  LY]+\lambda \mu[ L X, Y]+ \mu^2 [X,Y]=$$ $$=\lambda L[X,Y]_{\lambda L}+\mu[X,Y]_{\lambda L}+\lambda \mu L[X,Y]+\mu^2[X,Y],$$
from which the claim follows. 
\endproof

\begin{definition}
We call $(\Omega^*(M, TM), \dna, \dlna)$ the {\em hydrodynamic bidifferential complex}. 
\end{definition}

Let us present some examples of computations using the differentials $\dna, \dlna$. 
\begin{example}
Consider a 1-form with values in $TM$, $V^i_j$. 
Then 
$$(\dna V)^i_{jk}=\nabla_j V^i_k-\nabla_k V^i_j=\d_j V^i_k+\Gamma^i_{mj}V^m_k-\Gamma^m_{kj}V^i_m-\d_k V^i_j-\Gamma^i_{mk}V^m_j+\Gamma^m_{jk}V^i_m,$$
and thus
$$(\dna V)^i_{jk}=\d_j V^i_k+\Gamma^i_{mj}V^m_k-\d_k V^i_j-\Gamma^i_{mk}V^m_j.$$
Notice that from the definition we should write $\nabla_k V^i_{(j)}$ instead of $\nabla_k V^i_j$ meaning that
 computing covariant derivative the lower indices must be neglected or thought as frozen. However as the above computation shows
 the additional terms involving covariant derivatives of lower indices automatically cancel out.
\end{example}

%\begin{example}
%Let us consider $(\dna \delta)^i_{jk}$ where $\delta$ is the Kronecker delta, i.e. identity tensor of type (1,1).
%$$(\dna \delta)^i_{jk}=\na_j \delta^i_k-\na_k \delta^i_j=\Gamma^i_{mj}\delta^m_k-\Gamma^i_{mk}\delta^m_j=T^i_{kj}=0,$$
%the torsion of $\na$. 
%\end{example}

%\begin{example}
% Let $\omega^i_{jk}$ a $2$ form valued in $TM$. 
%Then 
%$$(\dna \omega)^l_{kij}=\d_k \omega^l_{ij}+\Gamma^l_{mk}\omega^m_{ij}-\d_i\omega^l_{kj}-\Gamma^l_{mi}\omega^m_{kj}+\d_j \omega^l_{ki}+   \Gamma^l_{mj}\omega^m_{ki}.$$
%\end{example}

%\begin{example}
 %Let $L^i_{jkl}$ be a three form with value in $TM$. Then 
%$$(\dna L)^i_{jklm}=\na_j L^i_{klm}-\na_k L^i_{jlm}+\na_l L^i_{jkm}-\na_m L^i_{jkl}.$$
%\end{example}
\begin{example}
Let us apply now $\dlna$ on a $1$-form with values in $TM$, $V^i_j$. 
We have $$(\dlna V)^i_{jk}=L^m_j(\na_m V^i_{(k)})-L^m_k(\na_m V^i_{(j)})-V^i_l(\d_j L^l_k-\d_k L^l_j).$$
Observe that in this case, due to the presence of $L$, we cannot forget that the covariant derivative involves only the upper
 indices.
\end{example}

Using $\dlna$ we can reformulate the condition for $L$ to have vanishing torsion in the following way:
\begin{proposition}
Let $\nabla$ be any torsionless connection of $TM$, then the endomorphism $L$ has vanishing Nijenhuis torsion if and only if $\dlna L=0$. 
\end{proposition}
\proof
We have 
$$(\dlna L)^i_{jk}=L^m_j \nabla_m L^i_{(k)}-L^m_k \nabla_m L^i_{(j)}-L^i_l([\partial_j, \partial_k]_L)^l.$$
Since $\nabla_m L^i_{(k)}=\partial_m L^i_k+\Gamma^i_{lm}L^l_k$ and analogously for $\nabla_m L^i_{(j)}$, we obtain that
$$L^m_j \nabla_m L^i_{(k)}-L^m_k \nabla_m L^i_{(j)}=L^m_j \partial_m L^i_k-L^m_k\partial_m L^i_j+L^m_jL^l_k\Gamma^i_{lm}-L^m_k L^l_j\Gamma^i_{lm}=$$
$$=L^m_j \partial_m L^i_k-L^m_k\partial_m L^i_j.$$
On the other hand, $L^i_l([\partial_j, \partial_k]_L)^l=L^i_l(\partial_j L^l_k-\partial_k L^l_j)$. Therefore, $\dlna L=0$ if and only if 
$$L^m_j \partial_m L^i_k-L^m_k\partial_m L^i_j=L^i_l(\partial_j L^l_k-\partial_k L^l_j),$$ which is exactly the condition $[L\partial j, L\partial_k]=L([\partial_j, \partial_k]_L)$, namely \eqref{nijenhisalternative}. This proves the claim. 
\endproof

\section{Semi-Hamiltonian systems. Tsarev's theory}
Let 
$$\Gamma^i_{ij}(u^1,\dots,u^n),\qquad i\ne j,\,i,j=1,\dots,n$$
be a set of functions of $n$ variables $(u^1, \dots u^n)$ satisfying the conditions
\begin{equation}\label{sh}
\d_i\Gamma^k_{kj}-\Gamma^k_{kj}\Gamma^j_{ij}
+\Gamma^k_{ik}\Gamma^k_{kj}-\Gamma^k_{ik}\Gamma^i_{ij}=0,\qquad\forall i\ne j\ne k\ne i
\end{equation}
and, consequently also the conditions
\begin{equation}
\label{shder}
\partial_j\Gamma^i_{ik}=\partial_k\Gamma^i_{ij},\qquad\forall i\ne j\ne k\ne i.
\end{equation}
Consider the system
\begin{equation}
\label{sym} 
\f{\d_j v^i}{v^j-v^i}=\Gamma^i_{ij},
\end{equation}
for the unknown functions $(v^1,\dots,v^n)$ of the $n$ variables $(u^1, \dots u^n)$. According to the results of \cite{ts} the equations
 \eqref{sh} are the compatibility conditions of \eqref{sym}. This implies that if \eqref{sh} are identically satisfied, the system \eqref{sym} admits a general
 solution depending on $n$ arbitrary functions each depending on one variable.
 Consider now the diagonal systems of PDEs of hydrodynamic type
\begin{equation}
\label{hts}
u^i_t=v^i(u)u^i_x\qquad i=1,...,n.
\end{equation}
and
\begin{equation}
\label{hts2}
u^i_{\tau}=w^i(u)u^i_x\qquad i=1,...,n.
\end{equation}
 defined by two different solutions $(v^1,\dots,v^n)$ and  $(w^1,\dots,w^n)$ of \eqref{sym}. The condition
 of commutativity of the flows
$$u^i_{t\tau}=u^i_{\tau t},$$
namely
$$\f{\d_j v^i}{v^j-v^i}=\f{\d_j w^i}{w^j-w^i},$$
is clearly satisfied. In other words the solutions of the
 system \eqref{sym} define a family of commuting
 flows for diagonal systems of PDEs of hydrodynamic type.  With a little abuse of terminology we will call
 such a family a \emph{semihamiltonian hierarchy}.

To conclude this brief section we recall from \cite{ts} that the equations (\ref{sh}) are also the integrability conditions for the system
\begin{equation}
\label{cl}
\d_i\d_j H-\Gamma^i_{ij}\d_i H-\Gamma^j_{ji}\d_j H=0, \; i\neq j
\end{equation}
which provides   the densities $H$  of conservation laws for (\ref{hts}). 

\section{$F$-manifold with eventual identities.}
The notion of $F$-manifold was introduced in \cite{HM} as a generalization of the concept of Frobenius manifold. Let us recall that an  $F$-manifold is a manifold endowed with a $(1,2)$-tensor field $c$ satisfying the conditions
\begin{eqnarray}
\label{F1}
&&c^i_{jk}=c^i_{kj}\\
\label{F2}
&&c^i_{jl}c^l_{km}=c^i_{kl}c^l_{jm}\\
\label{F3}	
&&(\d_s c^k_{jl})c^s_{im}+(\d_j c^s_{im})c^k_{sl}-(\d_s c^k_{im})c^s_{jl}-(\d_i c^s_{jl})c^k_{sm}-(\d_l c^s_{jm})c^k_{si}-(\d_m c^s_{li})c^k_{js}=0\ .
\end{eqnarray}
The tensor $c$ induces a bilinear product on vector fields:
$$
(X\circ Y)^i:=c^i_{jk}X^j Y^k\ \quad \text{ for all vector fields } X, Y.
$$
Due to \eqref{F1}, \eqref{F2} and \eqref{F3} the product is commutative 
 associative and satisfies the Hertling-Manin condition
\beq
\label{hmcond}
\begin{aligned}
&&[X\circ Y,Z\circ W]-[X\circ Y,Z]\circ W-[X\circ Y,W]\circ Z-X\circ[Y,Z\circ W]+X\circ[Y,Z]\circ W+\\
&&+X\circ[Y,W]\circ Z-Y\circ[X,Z\circ W]+Y\circ[X,Z]\circ W+Y\circ[X,W]\circ Z=0\ .
\end{aligned}
\eeq
Usually in the definition of an $F$-manifold one also assumes the existence of a vector field $e$ called \emph{unit} which behaves like a unit for the product $\circ$, namely 
\beq\label{F0}
c^i_{jl}e^l=\delta^i_j
\eeq
or, equivalently: 
$$X\circ e=X,\qquad\forall X.$$

In many cases, one can introduce
 different  products satisfying conditions \eqref{F0}, \eqref{F1}, \eqref{F2} and \eqref{F3} on the same manifold. In the case of semisimple
  $F$-manifolds with compatible connection (see the next section) this freedom is related to the arbitrariness in the choice of Riemann invariants of the associated semihamiltonian hierarchy \cite{LPR}. More in general,
 this freedom is due to the existence of special vector
 fields called by Manin \cite{manin} \emph{eventual identities}. The most important 
 examples of geometric structures admitting eventual identities are the almost Frobenius manifolds introduced by Dubrovin in \cite{Dad}: indeed in this
 case the eventual identity is provided by the Euler vector field itself.

\begin{defi} 
A vector field E on an $F$-manifold is called an eventual identity, if it is invertible with respect to $\circ$ (i.e. there is a vector field $E^{-1}$ such that
$E\circ E^{-1} = E^{-1} \circ E = e$) and, moreover, the bilinear product $*$ defined via
\beq\label{nm}
X *Y := X \circ Y \circ E^{-1},\qquad \text{ for all } X, Y \text{ vector fields}
\eeq
defines a new F-manifold structure on M.
\end{defi}
The vector field $E$ is, by definition, the unit of the product $*$. This is the origin of the name  \emph{eventual identity}.
A characterization of eventual identities was recently given in \cite{DS}. 
\begin{theorem}\cite{DS}
An invertible vector field $E$ is an eventual identity if and only if
\beq\label{DScond}
{\rm Lie}_E(\circ)(X,Y)=[e,E]\circ X\circ Y,\qquad\forall X,Y.
\eeq
\end{theorem}
Special cases of eventual identities are the Euler vector 
 fields. In this case
\begin{eqnarray*}
&&{\rm Lie}_E(\circ)(X,Y)=X\circ Y\\
&&[e,E]=e
\end{eqnarray*}
Let us recall the following important definition: an $F$-manifold is called \emph{semisimple} if there exists a distinguished system of coordinates, called \emph{canonical coordinates} such that the tensor $c$ has the following form in these coordinates:
$$c^i_{jk}=\delta^i_j\delta^i_k.$$

\begin{theorem}\label{theoremDSNijenhuis}
Consider a semisimple $F$-manifold and assume that the eigenvalues of the endomorphism $V=E\,\circ$ are distinct. Then condition \eqref{DScond}
is equivalent to the vanishing of the Nijenhuis torsion of $V$.
\end{theorem}

\n
\emph{Proof}.  Suppose the Nijenhuis torsion of $V$ vanishes. This means that 
$$N_V(X,Y)=[VX,VY]-V\,[X,VY]-V\,[VX,Y]+V^2\,[X,Y]=0$$
for any pair $(X,Y)$ of vector fields. In local coordinates
 this means that
\begin{equation*}
\sum_{s=1}^n (V^s_i\d_s V^k_j-V^s_j\d_s V^k_i+V^k_s\d_j V^s_i
-V^k_s\d_i V^s_j)=0.
\end{equation*}
Since the $F$-manifold is assumed to be semisimple, there exists a system of coordinates  (canonical coordinates) in which $V$ is diagonal $V^i_j=c^i_{jk}E^k=E^i\delta^i_j$ and therefore the previous sum reads
\begin{equation*}
\delta^k_j(E^i-E^j)\d_i E^j -\delta^k_i(E^j-E^i)\d_j E^i=0.
\end{equation*}
If $E^i\ne E^j$ this is equivalent to
$$\d_j E^i=0,\qquad\forall j\ne i.$$

Now suppose that condition  \eqref{DScond} is satisfied. In local coordinates it reads
$$E^m\d_m c^i_{jk}+c^i_{lk}\d_j E^l+c^i_{lj}\d_k E^l-c^l_{jk}\d_l E^i=c^i_{lj}c^l_{km}[e,E]^m.$$
In canonical coordinates $e^i=1$ for all $i$ and therefore:
\beq\label{auxiliaryxxx}
\delta^i_k\d_j E^i+\delta^i_j\d_k E^i-\delta^j_k\d_j E^i=\delta^i_j\delta^i_k \delta^i_m \sum_s\d_s E^m.\eeq 
If all the indices in \eqref{auxiliaryxxx} are equal, then the equation is identically satisfies due to the fact that in canonical coordinates, in the semisimple case $E^i$ is just a function of the $i$-th coordinate. In the case when two indices are equal and the third is different we have: 

\n
- $k=j\ne i$. In this case $-\d_j E^i=0$,

\n
- $k=i\ne j$. In this case $\d_j E^k=0$,

\n
- $i=j\ne k$. In this case $\d_k E^i=0$.

These are exactly the vanishing Nijenhuis torsion conditions as we have seen above.  
Finally, if all the indices $i,j,k$ are distinct, equation \eqref{auxiliaryxxx} is automatically satisfied as it is immediate to see.
\endproof

From one hand the above theorem is quite surprising
 since the condition \eqref{DScond} is \emph{linear} in $E$ while the Nijenhuis condition is  quadratic in $E$:
$$N_{V}(X,Y)=[E\circ X,E\circ Y]+E\circ E\circ [X,Y]-E\circ [X,E\circ Y]-E\circ [E\circ X,Y].$$
 From the other side is not surprising since, as we mentioned above, in the case
 of $F$-manifolds with compatible connection the freedom in the choice of the eventual identities is related to the reparametrization of the Riemann invariants. Let us consider now the general case.
 A relation between \eqref{DScond} and the Nijenhuis
 condition is still present. However it goes only in one direction.
 
\begin{theorem}
If $E$ satisfies  \eqref{DScond}, in particular if $E$ is an eventual identity, then the Nijenhuis torsion of $V=E\,\circ$ vanishes.
\end{theorem}

\n
\proof 
First we rewrite the condition for $E$ to be an eventual identity in a different way. Since $X\circ Y$ can be viewed as the complete contraction of the tensor field $c$ with the vector fields $X$ and $Y$ and since the Lie derivative commute with any contraction, we have that 
$${\rm Lie}_E(\circ)(X,Y)={\rm Lie}_E(\circ(X,Y))-\circ({\rm Lie}_E X, Y)-\circ(X, {\rm Lie}_EY)=$$
$$=[E, X\circ Y]-X\circ[E, Y]-[E, X]\circ Y.$$
In this way, \eqref{DScond} can be written as 
\beq\label{DScond2}
[E, X\circ Y]+X\circ[Y, E]+[X, E]\circ Y=[e, E]\circ X\circ Y.
\eeq

Now we proceed to write the 
 Nijenhuis condition $[E\circ X,E\circ Y]+E\circ E\circ [X,Y]-E\circ [X,E\circ Y]-E\circ [E\circ X,Y]=0$
 in a different form. Expanding $[E\circ X,E\circ Y]$ using the Hertling-Manin condition (\ref{hmcond}) and the properties of $\circ$, the Nijenhuis condition can be rewritten as
\beq\label{Nij2}
N_{V}(X,Y)=[X\circ E,E]\circ Y-[X,E]\circ E\circ Y+[E,Y\circ E]\circ X-[E,Y]\circ X\circ E=0.
\eeq
Specializing the previous expression with $Y=e$ we get:
\beq\label{Nij3}
[X\circ E,E]-[X,E]\circ E-[E,e]\circ X\circ E=0
\eeq
Surprisingly \eqref{Nij3} implies \eqref{Nij2}. Indeed
\begin{eqnarray*}
&& N_V(X,Y)=\\
&&Y\circ([X\circ E,E]-[X,E]\circ E)-X\circ([Y\circ E,E]-[Y,E]\circ E)=\\
&&Y\circ[E,e]\circ X\circ E-X\circ[E,e]\circ Y\circ E=0.
\end{eqnarray*}
This means that vanishing of the Nijenhuis torsion
 is equivalent to  \eqref{Nij3}. 
In order to prove the theorem it is sufficient to observe that \eqref{DScond2} reduces to \eqref{Nij3} when $Y=E$. Therefore for any vector field $E$ which is an eventual identity, we have that $E\,\circ$ has zero Nijenhuis torsion. 
 \endproof
 
Let us remark that once we write \eqref{DScond2} as $Q_E(X,Y):=[E, X\circ Y]+X\circ[Y, E]+[X, E]\circ Y-[e, E]\circ X\circ Y$, this appears manifestly as a bilinear symmetric form. Therefore by polarization identity it can be written as $Q_E(X,Y)=\frac{1}{2}Q_E(X+Y,X+Y)-\frac{1}{2}Q_E(X-Y,X-Y)$. This implies that for \eqref{DScond} is equivalent to  
$${\rm Lie}_E(\circ)(Z,Z)=[e,E]\circ Z^2, \; \text{ for all } Z $$
or 
$${\rm Lie}_E(Z\circ Z)=2Z\circ {\rm Lie}_E (Z)-{\rm Lie}_E(e)\circ Z^2.$$
\section{$F$-manifolds with compatible connection}
Let us introduce a special class of $F$-manifolds \cite{HM}.
\begin{defi}
\label{defi:fmancc}
An \emph{$F$-manifold with compatible connection} \cite{LPR} is a manifold endowed with an associative commutative multiplicative structure given by a $(1,2)$-tensor field $c$ and a torsionless connection $\nabla$ satisfying the following conditions 
\beq
\label{scc}
\nabla_l c^i_{jk}=\nabla_j c^i_{lk}
\eeq
and
\beq
\label{shc}
R^k_{lmi}c^n_{pk}+R^k_{lip}c^n_{mk}
+R^k_{lpm}c^n_{ik}=0.
\eeq
where
$$
R^i_{jkh}=\d_k\Gamma^i_{hj}-\d_h\Gamma^i_{kj}-
\Gamma^i_{hl}\Gamma^l_{kj}
+\Gamma^i_{kl}\Gamma^l_{hj}$$
are the components of the Riemann tensor.
\end{defi}
Notice that in the previous definition we did not impose explicitly the requirement that $c$ satisfies the Hertling-Manin condition; this is due to the fact that if the product is symmetric and associative and $c$ satisfies equation \eqref{scc}, then the Hertling-Manin condition is automatically fulfilled (this is proved in \cite{He}).

Let us observe also that in the interesting paper \cite{DS2}, the compatibility of 
the connection with the product is intended  in a weaker sense, since no 
  restrictions are imposed there on the Riemann tensor. 
  
Conditions \eqref{scc} and \eqref{shc} can also be 
 written respectively as
\beq\label{sccinv}
\left(\nabla_X \circ\right)\left(Y,Z\right)=\left(\nabla_Y \circ\right)\left(X,Z\right),
\eeq
and
\beq\label{rc-intri}
Z\circ R(W,Y)(X)+
W\circ R(Y,Z)(X)+Y\circ R(Z,W)(X)=0,
\eeq
for any choice of the vector fields $(X,Y,W,Z)$.

When studying the systems of PDEs that control the densities of conservation laws, an alternative form of \eqref{rc-intri}, equivalently \eqref{shc} will be handy. These are provided by the following 
\begin{lemma}
Let $M$ be an $F$-manifold with compatible connection $\nabla$. Then \eqref{rc-intri} is equivalent to 
\beq\label{rc-intri2}
R(Y,Z)(X\circ W)+R(X,Y)(Z\circ W)+R(Z,X)(Y\circ W)=0,
\eeq
for every vector fields $X,Y,W,Z$. Moreover, the above condition in local coordinates reads
\beq\label{shc2}
R^n_{kmp}c^k_{il}+R^n_{kim}c^k_{pl}+R^n_{kpi}c^k_{ml}=0.
\eeq
\end{lemma}
\proof
Consider the deformed connection ${\tilde \nabla}_X Y:=\nabla_X Y+z\, X\circ Y, \, z\in \mathbb{C}$, depending on the parameter $z$. Due to associativity of the product $\circ$ and the symmetry condition \eqref{sccinv}, the curvature tensor of this connection does not depend on $z$ (see \cite{St}). Now the second Bianchi identity gives
$$0={\tilde \nabla}_XR(Y,Z)(W)+{ \tilde \nabla}_ZR(X,Y)(W)+{\tilde \nabla}_YR(Z,X)(W)=$$
$$=X\circ R(Y,Z)(W)+Z\circ R(X,Y)(W)+Y\circ R(Z,X)(W)$$
$$-R(Y,Z)(X\circ W)-R(X,Y)(Z\circ W)-R(Z,X)(Y\circ W).$$
Since $X\circ R(Y,Z)(W)+Z\circ R(X,Y)(W)+Y\circ R(Z,X)(W)=0$ is \eqref{rc-intri}, one sees immediately the equivalence of \eqref{rc-intri2} and \eqref{rc-intri}. The coordinate expression is immediate. 
\endproof

For a semisimple $F$-manifold, in canonical coordinates, condition \eqref{scc} reads
\begin{eqnarray}
\label{scc3}
\Gamma^i_{jj}&=&-\Gamma^i_{ji}\\
\label{scc4}
\Gamma^i_{jk}&=&0,\qquad\forall i\ne j\ne k\ne i
\end{eqnarray}
and condition \eqref{shc} is equivalent to
\begin{eqnarray}
\label{shc3}
R^i_{ijk}&=&\d_j\Gamma^i_{ki}-\d_k\Gamma^i_{ji}=0,\qquad\forall i\ne j\ne k\ne i,\\
\label{shc4}
R^i_{jjk}&=&-\d_k\Gamma^i_{jj}-
\Gamma^i_{ki}\Gamma^i_{jj}
-\Gamma^i_{kk}\Gamma^k_{jj}
+\Gamma^i_{jj}\Gamma^j_{jk}=0,\qquad\forall i\ne j\ne k\ne i
\end{eqnarray}

\begin{remark}
If $u\to\tilde u$, the Christoffel symbols transform as
$$\tilde\Gamma^k_{ij}=\f{\d\tilde u^k}{\d u^r}
\left(\Gamma^r_{pq}\f{\d u^p}{\d\tilde u^i}\f{\d u^q}{\d\tilde u^j}+\f{\d^2 u^r}{\d\tilde u^i\d\tilde u^j}\right).$$
This means that if condition \eqref{scc4} is satisfied in canonical coordinates, then it is satisfied
 in any coordinates system $(\tilde{u}^1,\dots,\tilde{u}^n)$ related to canonical coordinates by a change of variables of the form $\tilde{u}^i=\tilde{u}^i(u^i),\,i=1,\dots,n$. This will be instrumental in defining the notion of almost hydrodynamically equivalent connections. 
\end{remark}
\begin{remark}
Compatible connections are not uniquely defined. For instance, in canonical coordinates,
 if the Christoffel symbols $\Gamma^i_{jk}$ define a compatible connection, then 
 the Christoffel symbols
$\tilde\Gamma^i_{jk}$ with 
$$\tilde\Gamma^i_{jk}:=\Gamma^i_{jk},\qquad i\ne j\,{\rm or}\,i\ne k,\,{\rm or}\,j\ne k, \; \text{ and } \; \tilde\Gamma^i_{ii} \; \text{ arbitrary}$$
define also a compatible connection \cite{LP}.
\end{remark}

Additional structures or requirements might be added to an $F$-manifold with compatible connection. Among them let us mention the most relevant ones for our investigation:
\begin{itemize}
\item The existence of a unit vector field $e$:
\beq\label{unitycond}
c^i_{jk}e^k=\delta^i_j.
\eeq
\item The existence of an eventual identity $E$. 

In the semisimple case, in canonical coordinates we have therefore 
$$e^i=1,\qquad E^i=E^i(u^i),\qquad L^i_j=E^j\delta^i_j,$$
where $L$ is the endomorphism of the tangent bundle given by $L:=E\circ$. 
\item The flatness of the connection $\nabla$ \cite{manin}. In this
 case starting from a frame of flat vector fields it is 
 possible to define a hierarchy of quasilinear PDEs of the form
$$u^i_t=V^i_j(u)\,u^j_x=c^i_{jk}(u)X^k\,u^j_x$$
called \emph{principal hierarchy}. In the case of Frobenius manifolds, this hierarchy was introduced by Dubrovin \cite{du93}. The straightforward generalization under the weaker Manin's assumptions was given in \cite{LPR}.
\item The existence of an  invariant metric $\eta$ satisfying 
$$\nabla\eta=0.$$
In this case, the $F$-manifold is called  \emph{Riemannian $F$} \cite{LPR}.  
In the flat case the invariant metric defines a local Poisson structure for
 the principal hierarchy. This means that the equations
 of the hierachy can be written in the form
\beq\label{hamiltonian}
u^i_t=V^i_j(u)\,u^j_x=P^{ij}\f{\d H}{\d u^j},\qquad
i=1,\dots,n
\eeq 
for a suitable local functional $H[u]$. The differential operator
\beq\label{DN}
P^{ij}=\eta^{ij}\d_x-\eta^{il}\Gamma^j_{lk}u^k_x
\eeq
is the local Poisson bivector of hydrodynamic type associated to the flat metric
 $\eta$ \cite{dn89}. 
\item The existence of a second flat connection compatible with the multiplicative structure defined 
 by the eventual identity $E$. In the case of Frobenius manifolds such a connection is the Levi-Civita 
 connection of the 
 \emph{intersection form} $g$.  The pencil $g_{\lambda}=g-\lambda\eta$ is a \emph{flat pencil of metrics} \cite{du97}. In the case of Frobenius manifolds, the multiplication by the Euler vector field $L=E\circ$
 is related to the flat pencil $g_{\lambda}$ by:
$$L^i_j=\eta_ {jl}g^{li}.$$
\end{itemize}
This list might be extended including the additional axioms appearing in the definition of a Frobenius manifold. Since these additional assumptions will not be used in the paper we refer the reader to the literature (for instance \cite{du93}) for more details.

%From what we said above it follows that on \emph{a $F$-manifold with compatble flat connection $\nabla$ and eventual identity $E$ one can define the bidifferential complex 
%$$(\Omega^*(M, TM), \dna, \dlna),$$
%where $L=E\circ$.}

\vspace{.5 cm}
\n
{\bf Assumption 1}: \emph{From now on we will deal only with \emph{semisimple} $F$-manifolds}.
\vspace{.2 cm}

\n
{\bf Assumption 2}: \emph{From now on we will
 assume that the components of an eventual identity $E$ in canonical coordinates are functionally independent.}
\bigskip

\noindent Under these two assumptions we can prove the following results upon which all our subsequent investigations are based:

\begin{theorem}\label{fundamentaltheorem}
Assume assumptions $1$ and $2$ above hold. Then we have
\begin{enumerate} 
\item The canonical coordinates  $(\tilde{u}^1,\dots,\tilde{u}^n)$ for the product $*$ and  the canonical coordinates $(u^1,\dots,u^n)$ for the product $\circ$ are related through a simple reparametrization of the form  $\tilde{u}^i=\tilde{u}^i(u^i),\,i=1,\dots,n.$
\item The endomorphism $L:=E\, \circ$ can be written as $L=\tilde E\, *$; moreover  $\tilde E=E\circ E$ and $\tilde E^i\neq \tilde E^j$ for $i\neq j$ where $\tilde E^i$ are the components of $\tilde E$ in canonical coordinates for $*$. 
\end{enumerate}
\end{theorem}
\proof
Proof of the first claim: Since by assumption $1$ the $F$-manifold is semisimple, and $E$ is an eventual identity, then by Theorem \ref{theoremDSNijenhuis},
 $L=E\,\circ$ has vanishing Nijenhuis torsion. This implies that in canonical coordinates for $\circ$ we have $E^i=E^i(u^i)$. By assumption $2$ the components $E^i$ are all functionally independent. 
By definition (compare equation \eqref{nm}) the product $*$ in canonical coordinates for $\circ$ reads
$$c^{*i}_{jk}=\f{1}{E^i}\delta^i_j\delta^i_k.$$
Now we look for a reparametrization of the canonical coordinates $u^i$ of the form $u^i\mapsto \tilde u^i(u^i)$ in such a way that in these new coordinates the constant structures for $*$ appear to have the form $\delta^i_j\delta^i_k$. This will prove that the canonical coordinates for $*$ are indeed obtained via a simple reparametrization of the canonical coordinates for $\circ$. 

Indeed, we have $$c^{*i}_{jk}\to\tilde{c}^{*l}_{mn}=c^{*i}_{jk}\f{\d\tilde{u}^l}{\d u^i}\f{\d u^j}{\d \tilde{u}^m}\f{\d u^k}{\d 
\tilde{u}^n}=\f{1}{E^i}\delta^i_j\delta^i_k \f{\d\tilde{u}^l}{\d u^i}\f{\d u^j}{\d \tilde{u}^m}\f{\d u^k}{\d 
\tilde{u}^n}=$$     $$=\f{1}{E^i}\f{\d\tilde{u}^l}{\d u^i}\f{\d u^i}{\d \tilde{u}^m}\f{\d u^i}{\d 
\tilde{u}^n}=\f{1}{E^l}\f{\d u^l}{\d \tilde{u}^l}\delta^l_m\delta^l_n=\f{1}{\tilde{E}^l}\delta^l_m\delta^l_n$$
where $\tilde{E}^l$ are the components of $E$ in the new coordinates $\tilde u^i$. 

\noindent If we choose the reparametrization to be given by: \beq\label{reparametrization1}\f{d\tilde{u}^l}{d u^l}=\f{1}{E^l(u^l)}\eeq
that is if $$\tilde{u}^l=\int \f{1}{E^l(u^l)}\,du^l,$$ then $\tilde{E}^l=\f{d \tilde{u}^l}{d u^l}E^l=1$ and 
$\tilde{c}^{*l}_{mn}=\delta^l_m\delta^l_n$. This proves the first claim and provides also an explicit form (equation \eqref{reparametrization1}) for the change of coordinates from one system of canonical coordinates to the other one.

Proof of the second claim: We look for a vector field $\tilde E$ such that $\tilde E* Y=E\circ Y$ for all vector fields $Y$. First of all notice that, even without assumptions $1$ and $2$, one has  
$E\circ Y=\tilde E * Y=\tilde E \circ Y \circ E^{-1}$ and therefore 
$$\tilde E=E\circ E,$$
simply because $E$ is an eventual identity. Let us remark that this $\tilde E$ has nothing to do with the one that appears in the proof of the first claim. 
 
%In cononical coordinates of $*$ $\tilde{E}^l=1$.
In canonical coordinates of $\circ$ 
we have that the components of the vector field $E\circ E$ are given by $(E\circ E)^i=(E^i)^2$. In canonical coordinates 
for $*$ one gets:
$$(\widetilde{E\circ E})^i(\tilde u)=(E^i)^2(u^i)\f{d\tilde u^i}{d u^i}=E^i(u^i(\tilde u^i))=\f{d u^i}{d\tilde u^i}$$

Clearly $\f{d u^i}{d \tilde{u}^i}$ is not constant, otherwise $\f{d\tilde u^i}{ d u^i}$ would be constant too and, because of equation \eqref{reparametrization1},  that would imply that $E^i(u^i)$ is constant, not depending on $u^i$, contrary to assumption 2.  Therefore we have 
 $$(\widetilde{E\circ E})^i\ne(\widetilde{E\circ E})^j,$$
since the right and the left hand side depend on different 
coordinates. This proves the second claim. 
\endproof

\section{Hydrodynamically equivalent and almost hydrodynamically equivalent connections}\label{equivalentsection}
Given a semisimple $F$-manifold with compatible connection one can define a semihamiltonian hierarchy.
 In canonical coordinates the functions  $\Gamma^i_{ij}$ $(i\ne j)$ defining the hierarchy according to \eqref{sym} 
 are a subset of the Christoffel symbols of the compatible connection (we will give later a coordinate free definition). The fact that the hierarchy is semihamiltonian follows immediately from \eqref{shc3}.

As we mentioned above, on the same $F$-manifold one can define different compatible connections and different multiplicative structures.

Let $(M,\circ,E)$ be a semisimple $F$-manifold with eventual identity $E$. We introduce the following definitions:

\begin{definition}
 Let $\nabla$ be a connection compatible with
 $\circ$ and $\tilde\nabla$ be a connection compatible with $*$. They are called \emph{almost hydrodynamically equivalent} if 
\beq\label{almostcomp}
(d_{\nabla}-d_{\tilde\nabla})(X\,\circ)=0,\qquad{\rm or}\qquad(d_{\nabla}-d_{\tilde\nabla})(X\,*)=0
\eeq
 for every vector field $X$.% and $\tilde X$ such that $X\, \circ=\tilde X\, *$. Notice that $\tilde X=E\circ X$ as it is immediate to check. 

\noindent In canonical coordinates for $\circ$ the equation \eqref{almostcomp} (on the left) reads
\beq\label{almostcomp2}
(X^k-X^j)(\Gamma^i_{kj}-\tilde\Gamma^{i}_{kj})=0.\eeq
where $\tilde\Gamma^i_{jk}$ are the Christoffel symbols of the connection $\tilde \nabla$ {\em in the canonical coordinates of $\circ$}. Similarly for the equation on the right using canonical coordinates for $*$.
\end{definition}

\noindent To see why \eqref{almostcomp} gives \eqref{almostcomp2}, observe that
in canonical coordinates for the product $\circ$ the Christoffel symbols $\Gamma^i_{kj}$
 with $i\ne j\ne k\ne i$ vanish, because of the compatibility of  
$\nabla$ with $\circ$. Now due to the compatibility of $\tilde\nabla$ with $*$ one has that $\tilde\Gamma^r_{pq}=0$ for $r\neq p\neq q\neq r$ in the canonical coordinates for $*$. On the other hand, in the canonical coordinates for $\circ$ one has 
$$\tilde\Gamma^{k}_{ij}=\f{\d u^k}{\d \tilde u^r}\f{ \d \tilde u^p}{\d u^i}\f{\d \tilde u^q}{\d u^j}\tilde\Gamma^r_{pq}+\f{\d u^k}{\d \tilde u^m}\f {\d^2 \tilde u^m}{\d u^i \d u^j},$$
where $\tilde\Gamma^r_{pq}$ are expressed in the canonical coordinates of $*$, while $\tilde\Gamma^{k}_{ij}$ are in the canonical coordinates of $\circ$.
Using the first claim of Theorem \ref{fundamentaltheorem}, one finds $\tilde\Gamma^{k}_{ij}=0$ for $i\ne j\ne k\ne i$. 
The same reasoning also shows that the two equations $(d_{\nabla}-d_{\tilde\nabla})(X\,\circ)=0$ and $(d_{\nabla}-d_{\tilde\nabla})(X\,*)=0$ are indeed equivalent. 

This means that the almost hydrodynamical equivalence of two connections
 is equivalent, in canonical coordinates $(u^1,\dots,u^n)$ or in any coordinates system 
 related to them by a reparametrization of the form $u^i\mapsto \tilde u^i(u^i)$, to the condition
$$\tilde\Gamma^{i}_{ij}=\Gamma^i_{ij},\qquad i\ne j.$$
For this reason \emph{almost hydrodynamically equivalent connections define the same semihamiltonian hierarchy}.
\begin{remark}\label{importantremark} For future use, it is important to notice that if $\nabla$ and $\tilde \nabla$ are two hydrodynamically almost equivalent connections, then for any endomorphism $V$ of $TM$ with the property that $V$ is diagonal both in the canonical coordinates for $\circ$ and in the canonical coordinates for $*$, we have that 
\beq\label{auxiliaryxxxx} (d_{\nabla}-d_{\tilde \nabla}) V=0. \eeq This follows directly from the definition.  \end{remark}

\begin{definition}
 Two connections $\nabla$ and $\tilde\nabla$  compatible with the same product $\circ$
 and almost hydrodynamically equivalent are called \emph{hydrodynamically equivalent}.  
 In concrete terms, in canonical coordinates, this means that
$$\tilde\Gamma^i_{jk}=\Gamma^i_{jk},\qquad i\ne j\,{\rm or}\,i\ne k,\,{\rm or}\,j\ne k.$$
\end{definition}

%\begin{bf}
\begin{remark}
The diagonal  metrics $g$  defining the local Hamiltonian structures of a given  
semi-Hamiltonian system
$$u^i_t=v^i(u) u^i_x,\qquad i=1,\dots,n,$$
are the flat solutions (if they exist) of the linear system of PDEs \cite{dn89}
$$\d_j\ln{\sqrt{g_{ii}}}=\f{\d_j v^i}{v^j-v^i}.$$
Notice that the Levi-Civita connections associated with  the solutions of the above system are automatically almost hydrodynamically equivalent. Indeed
$$\d_j\ln{\sqrt{g_{ii}}}=\Gamma^i_{ij}.$$
\end{remark} 
%\end{bf}

\section{Conservation laws}
In this section we will study the conservation laws of
 the semihamiltonian hierarchy associated to a semisimple $F$-manifold with compatible connection $\nabla$ and eventual identity $E$.

First of all, we observe the following, 
\begin{proposition} The linear system of PDEs for densities of conservation laws can be written in intrinsic form as
\beq\label{claws2}
(d_Ld h)+\mathcal{C}(d_{\nabla} L\otimes dh)=0.
\eeq
where $L=E\,\circ$ and $\mathcal{C}$ is the contraction of the tangent bundle-valued two-form $d_{\nabla} L$ with the one form $dh$. 
\end{proposition}
\proof  Taking into account \eqref{scc4}, in canonical coordinates the equation \eqref{claws2} reads
\beq\label{claws3}
(E^k-E^j)(\d_j\d_k h-\Gamma^j_{jk}\d_j h-\Gamma^k_{jk}\d_k h)=0
\eeq
that is clearly equivalent to \eqref{cl}. Notice that the
 equation \eqref{claws3} does not change if we substitute $\nabla$ with a hydrodynamically equivalent or an almost hydrodynamically equivalent connection. This motivated the introduction of the previous definitions.
 \endproof

\begin{remark}
The system \eqref{claws2} is completely characterized by the tensor field $d_{\nabla} L$.
In the case of Frobenius manifolds, in flat coordinates the components of the Euler vector
 field are linear functions:  
$$(\nabla E)^i_j=q^i\delta^i_j,$$
for constants $q^i$. 
This means that, in such coordinates (if $q^i\ne q^j$!) the structure constants can be  written in terms of $d_{\nabla}L$:
$$c^i_{jk}=\f{\d_j L^i_k-\d_k L^i_j}{q^j-q^k}=\f{(d_{\nabla}L)^i_{jk}}{q^j-q^k}.$$
\end{remark}

\begin{proposition}
The compatibility conditions for equation $(d_Ld h)+\mathcal{C}(d_{\nabla} L\otimes dh)=0$ follow from condition \eqref{rc-intri2}
$$R(Y,Z)(X\circ W)+R(X,Y)(Z\circ W)+R(Z,X)(Y\circ W)=0,$$
and condition \eqref{sccinv}
or equivalently, in coordinates from conditions \eqref{scc3}, \eqref{scc4}, \eqref{shc3} and \eqref{shc4}.
\end{proposition}
\proof
From the form of equation \eqref{claws2}, it follows that the compatibility conditions are that the two-form $(\mathcal{C}(d_{\nabla} L\otimes dh))_{jk}=(dh)_l\,(d_{\nabla} L)^l_{jk}$ must be exact with respect to both differential $d$ and $d_L$ for any density of conservation law $h$.
 First we compute $(d\mathcal{C}(d_{\nabla} L\otimes dh)))_{ijk}$. To simplify notation, call $S_{ij}:=(\mathcal{C}(d_{\nabla} L\otimes dh))_{ij}$. Then one has
 $$(dS)_{ijk}=\d_i(S_{jk})-\d_j (S_{ik})+\d_k (S_{ij}).$$
 Therefore, since \beq\label{auxiliaryx}\d_i(S_{jk})=(\nabla_i (dh)_l)(d_{\nabla}L)^l_{jk}+dh_l \nabla_i (d_{\nabla}L)^l_{jk},\eeq one has
 $$(dS)_{ijk}= (\nabla_i dh_l)(d_{\nabla}L)^l_{jk}-(\nabla_j dh_l)(d_{\nabla}L)^l_{ik}+(\nabla_k dh_l)(d_{\nabla}L)^l_{ij}+$$
 $$+dh_l\left(\nabla_i (d_{\nabla} L)^l_{jk}-\nabla_j (d_{\nabla} L)^l_{ik}+\nabla_k (d_{\nabla} L)^l_{ij} \right).$$
 In canonical coordinate, for a semisimple $F$-manifold with compatible connection we have that $\nabla_i (dh)_j=\d_i \d_j h-\Gamma^i_{ij}\d_i h -\Gamma^{j}_{ji}\d_j h$, for $i\neq j$. On the other hand, this is just the equation for the densities of conservation laws, namely equation \eqref{cl}, and therefore $\nabla_i (dh)_j=0$ for $i\neq j$, since $h$ is supposed to be a density of conservation laws.
 
 This means that $ (\nabla_i dh_l)(d_{\nabla}L)^l_{jk}-(\nabla_j dh_l)(d_{\nabla}L)^l_{ik}+(\nabla_k dh_l)(d_{\nabla}L)^l_{ij}$ is reduced to 
 \beq\label{remaining} (\nabla_i dh_i)(d_{\nabla}L)^i_{jk}-(\nabla_j dh_j)(d_{\nabla}L)^j_{ik}+(\nabla_k dh_k)(d_{\nabla}L)^k_{ij},\eeq
 with no sum over equal indices. 
 
 Since $L$ is diagonal in canonical coordinates, it is immediate to check that the previous expression is identically zero for $i\neq j\neq k\neq i$. This is also the case when $i=j=k$. It is also easy to check, again using the aforementioned property of $L$, that the \eqref{remaining} is identically zero when two of the indices are equal and the third is different. 
 
 This implies that $$(dS)_{ijk}=dh_l\left(\nabla_i (d_{\nabla} L)^l_{jk}-\nabla_j (d_{\nabla} L)^l_{ik}+\nabla_k (d_{\nabla} L)^l_{ij} \right)=dh_l\left(d^2_{\nabla} L \right)^l_{ijk},$$
 where the last equality on the right follows from the definition of $d_{\nabla}$. 
 
 Now expanding the expression for $\dna^2 L$ one gets
 $$[\dna^2 L]_{ijk}^l =\na_{[i}\na_{j]}L_k^l+\na_{[j}\na_{k]}L_i^l+\na_{[k}\na_{i]}L_j^l=R^l_{nij}L^n_k+R^l_{njk}L^n_i+R^l_{nki}L^n_j.$$
 At this point it is enough to observe that the last term on the right in the previous expression is just \eqref{shc2} saturated with $E$. Therefore if \eqref{shc2} is fulfilled, then $dS=0$ identically. (Obviously, if $\nabla$ is flat, condition \eqref{shc2} is fulfilled, and also $dS=0$ is satisfied since in this case $d_{\nabla}^2=0$.)
 
%ADESSO AGGIUNGERE CONTO CON $d_LS$ e far vedere che discende da \eqref{shc2}.

\noindent From the definition of $d_L$ one gets
$$(d_LS)_{ijk}=L^p_i \d_p S_{jk}-L^p_j \d_p S_{ik}+L^p_k\d_p S_{ij} $$
$$-S_{pi}(\d_jL^p_k-\d_k L^p_j)+S_{pj}(\d_iL^p_k-\d_k L^p_i)-S_{pk}(\d_iL^p_j-\d_j L^p_i).$$
Using \eqref{auxiliaryx}, we have that $L^p_i \d_p S_{jk}-L^p_j \d_p S_{ik}+L^p_k\d_p S_{ij}$ can be expanded to 
$$(dh)_l\left[ L^p_i \nabla_p (d_{\nabla}L)^l_{jk}-L^p_j \nabla_p (d_{\nabla}L)^l_{ik}+L^p_k\nabla_p (d_{\nabla}L)^l_{ij}\right]+$$
$$+(\nabla_p(dh)_l)\left[ L^p_i (d_{\nabla}L)^l_{jk}-L^p_j (d_{\nabla}L)^l_{ik}+L^p_k (d_{\nabla}L)^l_{ij}\right].$$
Since $h$ is a density of conservation laws, we have $\nabla_p(dh)_l=0$ for $p\neq l$. Therefore, in the the second line of the above expression, the only surviving terms corresponds to $l=p$.
Thus, the second line can be written as (sum over $p$):
$$ (\nabla_p(dh)_p)\left[ L^p_i (d_{\nabla}L)^p_{jk}-L^p_j (d_{\nabla}L)^p_{ik}+L^p_k (d_{\nabla}L)^p_{ij}\right],$$
and it is easy to check that in canonical coordinates this expression is identically vanishing, since $L$ is diagonal. 

Therefore $(d_LS)_{ijk}$ is reduced to 
$$(dh)_l\left[ L^p_i \nabla_p (d_{\nabla}L)^l_{jk}-L^p_j \nabla_p (d_{\nabla}L)^l_{ik}+L^p_k\nabla_p (d_{\nabla}L)^l_{ij}\right]$$
$$-S_{pi}(\d_jL^p_k-\d_k L^p_j)+S_{pj}(\d_iL^p_k-\d_k L^p_i)-S_{pk}(\d_iL^p_j-\d_j L^p_i),$$
or, 
$$(dh)_l\left[  L^p_i \nabla_p (d_{\nabla}L)^l_{jk}-L^p_j \nabla_p (d_{\nabla}L)^l_{ik}+L^p_k\nabla_p (d_{\nabla}L)^l_{ij}\right]$$
$$+(dh)_l\left[ -(\dna L)^l_{pi}(\d_jL^p_k-\d_k L^p_j)+(\dna L)^l_{pj}(\d_iL^p_k-\d_k L^p_i)-(\dna L)^l_{pk}(\d_iL^p_j-\d_j L^p_i)\right],$$
from which we recognize
$$(d_L S)_{ijk}=(dh)_l \left(d_{L\nabla} d_{\nabla}L \right)^l_{ijk}.$$
Observe that if $\nabla$ is flat and $L$ has zero Nijenhuis torsion, then $d_{L\nabla} d_{\nabla}L =-d_{\nabla}d_{L \nabla}L=0$ since $d_{L \nabla}L$ is precisely the Nijenhuis torsion of $L$.

Now we prove that \eqref{scc3}, \eqref{scc4}, \eqref{shc3} and \eqref{shc4} imply $d_{L\nabla}d_{\nabla}L=0$. First observe that $(d_{L\nabla}d_{\nabla}L)^l_{ijk}$ is automatically zero if at least two of the lower indices are equal, by skewsymmetry. Therefore, since we can assume $i\neq j\neq k\neq i$, in the expression for $(d_{L\nabla}d_{\nabla}L)^l_{ijk}$ the term 
$$\left[ -(\dna L)^l_{pi}(\d_jL^p_k-\d_k L^p_j)+(\dna L)^l_{pj}(\d_iL^p_k-\d_k L^p_i)-(\dna L)^l_{pk}(\d_iL^p_j-\d_j L^p_i)\right]$$
is identically vanishing, taking into account that in canonical coordinates $L^p_k=E^k(u^k)\delta^p_k$. For the same reason, the remaining expression for $(d_{L\nabla}d_{\nabla}L)^l_{ijk}$ in canonical coordinates is reduced to 
$$E^i \nabla_i (\nabla_j L^l_k-\nabla_k L^l_j)-E^j \nabla_j (\nabla_i L^l_k-\nabla_k L^l_i)+E^k\nabla_k (\nabla_i L^l_j-\nabla_j L^l_i).$$
Again, because of the special form of $L$ in canonical coordinates, we have that if the upper index $l$ is different from $i, j$ and $k$, then the previous expression is also identically vanishing. 

Without loss of generality, we can assume $l=i$, and in canonical coordinates we have 
\begin{equation}\label{conto}(d_{L\nabla}d_{\nabla}L)^i_{ijk}=E^j(\nabla_j \nabla_k E^i)-E^k(\nabla_k\nabla_j E^i),\end{equation}
with no summation on repeated indexes $j$ and $k$. Expanding the right hand side of equation \eqref{conto} and taking into account that in canonical coordinates the Christoffel symbols satisfy equations \eqref{scc3} and \eqref{scc4}, we obtain after long but straightforward computations
(no sum over repeated indices): 
$$(d_{L\nabla}d_{\nabla}L)^i_{ijk}=E^kE^j\left(\partial_k \Gamma^i_{ji}-\partial_j \Gamma^i_{ki}\right)+$$
$$+E^iE^j\left(\partial_j \Gamma^i_{ki}+\Gamma^i_{ji}\Gamma^i_{ki}-\Gamma^j_{jk}\Gamma^{i}_{ji}-\Gamma^k_{jk}\Gamma^i_{ki}\right)+$$
$$+E^kE^i\left(-\partial_k \Gamma^i_{ji}+\Gamma^j_{kj}\Gamma^i_{ji}-\Gamma^i_{ki}\Gamma^i_{ji}+\Gamma^k_{kj}\Gamma^i_{ki}\right).$$
Now the coefficient of $E^kE^j$ in the previous expression vanishes because of equation \eqref{shc3}, while the coefficients of $E^iE^j$ and $E^kE^i$ 
vanish due to equation \eqref{shc4}.
% Indeed, calling $S(h)=(dh)_l\,(d_{\nabla} L)^l_{jk}$, one has 
%\begin{eqnarray*}
%&&(dS)_{ijk}=(R^n_{ljk}L^l_{i}
%+R^n_{lij}L^l_{k}
%+R^n_{lki}L^l_{j})(dh)_n\\
%&&(d_L S)_{ijk}
%=(R^n_{klm}L^l_{i}L^m_{j}+R^n_{jml}L^l_{i}
%L^m_{k}+R^n_{ilm}L^l_{j}L^m_{k})(dh)_n\\
%\end{eqnarray*}
%Both conditions follows from \eqref{shc} (the second one in a non trivial way). They can be written in a nice form 
% in terms of $\dna$ and $d_{L\nabla}$.  Indeed
%$$[\dna^2 L]_{ijk}^l=[\dna (\dna L)](\d_i, \d_j, \d_k, dx^l)=\nabla_i((\dna L)_{jk}^l)-\na_j((\dna L)_{ki}^l)+\nabla_k((\dna L)_{ij}^l)=$$
%$$=\na_i(\na_j L_k^l-\na_jL_i^l)-\na_j(\na_iL_k^l-\na_kL_i^l)+\na_k(\na_i L_j^l-\na_j L_i^l)=$$
%$$=\na_{[i}\na_{j]}L_k^l+\na_{[j}\na_{k]}L_i^l+\na_{[k}\na_{i]}L_j^l=
%R^l_{nij}L^n_k+R^l_{njk}L^n_i+R^l_{nki}L^n_j.$$
%and AGGIUNGERE PARTE CON $d_L S$.

\endproof

\begin{example}
In the case of the $F$-manifold associated with the $\epsilon$-system \cite{Pv} it was shown \cite{LP}
 that 
$$\nabla_k L^i_j=(\nabla_k c^i_{jl})E^l+(1-n\epsilon)c^i_{kj}+\epsilon\d_k({\rm Tr} L)\,\delta^i_j$$
and therefore
$$(d_{\nabla} L)^i_{jk}=\epsilon\d_j({\rm Tr} L)\,\delta^i_k-
\epsilon\d_k({\rm Tr} L)\,\delta^i_j,$$
that implies
$$(d_Ld h)_{jk}=\left(dh\wedge d(\epsilon{\rm Tr}L)\right)_{jk}.$$
\end{example}

Let us consider now the recursion relations for densities of conservation laws in the case
 of $F$-manifolds with compatible flat connection and in the case of $F$-manifolds with a second flat connection compatible with the multiplicative structure $*$ defined by the eventual identity $E$.
We have the following
\begin{theorem}
In the case of an $F$-manifold with compatible flat connection, the densities of conservation laws obey the following recursion relations:
\beq\label{rrcl}
\nabla_j  (dh^{(p+1)})_i=c^l_{ij}(dh^{(p)})_l.
\eeq 
In the case of an $F$-manifold with compatible flat connection and possessing a second flat connection compatible with the multiplicative structure $*$ defined by the eventual identity $E$, the densities of conservation laws satisfy the following additional recursion relations:
\beq\label{rrcl2}
\nabla^{(1)}_h dh_k^{(\alpha+1)}=L^l_h\nabla^{(2)}_l dh_k^{(\alpha)},
\eeq
where $L=E\, \circ$.
\end{theorem}
\proof

In order to prove the above recurrence relations, we first observe that
\beq\label{ii}
L^m_h\nabla_m \omega_k-L^m_k\nabla_m \omega_h=
(d_{L\nabla}\omega)_{hk}+\omega_l(d_{\nabla}L)^l_{hk}.
\eeq
Indeed by definition
$$(d_{L\nabla}\omega)_{hk}
=(d_{L}\omega)_{hk}
=L^m_h\d_m\omega_k-L^m_k\d_m\omega_h-\omega_l
(\d_h L^l_k-\d_k L^l_h),$$
because $\omega$ is a scalar valued form,
while
$$\omega_l(d_{\nabla}L)^l_{hk}= \omega_l(\nabla_h L^l_k-\nabla_k L^l_h)=\omega_l(\d_h L^l_k-\d_k L^l_h
+\Gamma^l_{mh}L^m_k-\Gamma^l_{mk}L^m_h).$$
Combining the two expressions above one gets
\begin{eqnarray*}
&&(d_{L\nabla}\omega)_{hk}+\omega_l(d_{\nabla}L)^l_{hk}=\\
&&L^m_h\d_m\omega_k-L^m_k\d_m\omega_h+\omega_l
(\Gamma^l_{mh}L^m_k-\Gamma^l_{mk}L^m_h)=L^m_h\nabla_m \omega_k-L^m_k\nabla_m \omega_h.
\end{eqnarray*}
Now we show that if $h^{(p+1)}$ is related to $h^{(p)}$ via \eqref{rrcl}, then it also satisfies \eqref{claws2}.
Plugging in $h^{(p+1)}$ in \eqref {claws2} and using the identity \eqref{ii} we obtain
$$(d_{L\nabla}dh^{(p+1)})_{hk}+dh^{(p+1)}_l(d_{\nabla}L)^l_{hk}=L^l_h\nabla_l dh_k^{(p+1)}-L^l_k\nabla_l dh_h^{(p+1)}.$$
Assuming that $h^{(p+1)}$ satisfies the recursion relation \eqref{rrcl}, we have 
$$L^l_h\nabla_l dh_k^{(p+1)}-L^l_k\nabla_l dh_h^{(p+1)}=[c^l_{hm}c^j_{lk}-c^l_{km}c^j_{lh}]E^m dh_j^{(p)}=0,$$
due to the associativity of $\circ$. Therefore $h^{(p+1)}$ is a density of conservation laws, even if $h^{(p)}$ is not. 
Instead, to prove the other recurrence relations, we need to assume $h^{(\alpha)}$ is a solution of \eqref{claws2} with respect to the second connection. 

We proceed in a similar manner to prove \eqref{rrcl2}. 
Suppose $h^{(\alpha)}$ is a solution of the linear system of PDEs \eqref{claws2} determining the densities of conservation laws, namely 
$$(d_{L\nabla^{(2)}}dh^{(\alpha)})_{hk}+dh^{(\alpha)}_l(d_{\nabla^{(2)}}L)^l_{hk}=0.$$
By identity \eqref{ii}, we have that $dh^{(\alpha)}$ is a solution of the above equation if and only if 
\beq\label{auxiliaryii}
L^l_h\nabla^{(2)}_l dh_k^{(\alpha)}-L^l_k\nabla^{(2)}_l dh_h^{(\alpha)}=0
\eeq
Now we prove that $h^{(\alpha+1)}$ is a solution of \eqref{claws2} with respect to $\nabla^{(1)}$ if $h^{(\alpha)}$ is a solution of \eqref{claws2} with respect to $\nabla^{(2)}$ and $h^{(\alpha+1)}$ is related to $h^{(\alpha)}$ via \eqref{rrcl2}.
Indeed, using the identity \eqref{ii} we have
$$(d_{L\nabla^{(1)}}dh^{(\alpha+1)})_{hk}
+dh^{(\alpha+1)}_l(d_{\nabla^{(1)}}L)^l_{hk}=L^l_h\nabla^{(1)}_l dh_k^{(\alpha+1)}-L^l_k\nabla^{(1)}_l dh_h^{(\alpha+1)},$$
and plugging-in \eqref{rrcl2} on the right hand side we obtain
\beq\label{auxiliaryiii}
L^l_h\nabla^{(1)}_l dh_k^{(\alpha+1)}-L^l_k\nabla^{(1)}_l dh_h^{(\alpha+1)}=L^l_hL^m_l\nabla^{(2)}_m dh_k^{(\alpha)}-L^l_kL^m_l\nabla^{(2)}_m dh_h^{(\alpha)}.\eeq
Since $h^{(\alpha)}$ satisfies \eqref{auxiliaryii}, being a density of conservation laws with respect to $\nabla^{(2)}$, in particular it satisfies $L^m_l\nabla^{(2)}_m dh_k^{(\alpha)}-L^m_k\nabla^{(2)}_m dh_l^{(\alpha)}=0$. Using this equation on the right hand side of \eqref{auxiliaryiii} one finds
$$L^l_hL^m_l\nabla^{(2)}_m dh_k^{(\alpha)}-L^l_kL^m_l\nabla^{(2)}_m dh_h^{(\alpha)}=
L^l_hL^m_k\nabla^{(2)}_m dh_l^{(\alpha)}-L^l_kL^m_h\nabla^{(2)}_m dh_l^{(\alpha)}.$$
Finally, using 
%Using
%$$L^m_l\nabla^{(2)}_m dh_k^{(\alpha)}
%-L^m_k\nabla^{(2)}_m dh_l^{(\alpha)}$$
%and
$$\nabla^{(2)}_m dh_l^{(\alpha)}=\nabla^{(2)}_l dh_m^{(\alpha)}$$
on the right hand side of the last expression
we obtain
\begin{eqnarray*}
&&L^l_hL^m_k\nabla^{(2)}_m dh_l^{(\alpha)}-L^l_kL^m_h\nabla^{(2)}_m dh_l^{(\alpha)}=\\
&&L^l_hL^m_k\nabla^{(2)}_m dh_l^{(\alpha)}-L^l_kL^m_h\nabla^{(2)}_l dh_m^{(\alpha)}=\\
&&L^l_hL^m_k\nabla^{(2)}_m dh_l^{(\alpha)}-L^m_kL^l_h\nabla^{(2)}_m dh_l^{(\alpha)}=0.
\end{eqnarray*}
This proves the result about the second recurrence relation. 
\endproof
\begin{remark}
If the almost hydrodynamically equivalent connections are those associated with
 a flat pencil of metrics defining a bi-Hamiltonian structure, then the recurrence relations \eqref{rrcl2} can
 be written in the form 
\beq\label{rrcl3}
g_{(1)}^{jl}\nabla^{(1)}_l dh_k^{(\alpha+1)}=
g_{(2)}^{jl}\nabla^{(2)}_l dh_k^{(\alpha)}.
\eeq
They coincide with the usual Lenard-Magri recurrence relations
\beq\label{bih}
P_1^{ij}\f{\delta H^{(\alpha+1)}}{\delta u^j}=
P_2^{ij}\f{\delta H^{(\alpha)}}{\delta u^j}
\eeq
where $P_1$ and $P_2$ are the Poisson bivectors 
 of hydrodynamic type associated with $g_{(1)}$ and $g_{(2)}$ and 
$$H^{(\alpha)}=\int h^{(\alpha)}(u)\,dx,\qquad
H^{(\alpha+1)}=\int h^{(\alpha+1)}(u)\,dx.$$
However, we will see in the next few sections that in general one obtains recurrence relations that are more general compared to the usual ones coming from a Lenard-Magri chain. 
\end{remark}

%\begin{bf}
\begin{remark}
The equation for the densities of conservation laws can be also written in the form
\beq\label{clawsalt}
dd_V h=-d_Vd h=0
\eeq
where $V$ is one of tensor fields defining the symmetries. 
Indeed, in canonical coordinates $V^i_j=v^i\delta^i_j$ and therefore
\begin{eqnarray*}
(dd_V h)_{ij}&=&\d_i\left(v^j\d_j h\right)-\d_j\left(v^i\d_i h\right)=
(v^j-v^i)\d_i\d_j h+\d_i v^j\d_j h-\d_j v^i\d_i h\\
(d_Vd h)_{ij}&=&v^i\d_i\d_j h-v^j\d_j\d_i h+\d_i h\d_j v^i-\d_j h\d_i v^j=
-(dd_V h)_{ij}.
\end{eqnarray*}
%(MI CHIEDO DUNQUE SE L'ANTICOMMUTATIVITA' SEGUA DA HAANTJES NULLO)
 From \eqref{clawsalt} it follows that the 1-form $d_V h$ is (locally) exact: $d_V h=dk$. 
 The function $k$ is the current associated to $h$. Indeed in canonical coordinates $\d_i k=v^i\d_ih$ and this means that $\d_t h=\d_x k$. %AL MOMENTO NON MI VIENE IN MENTE
% NESSUNA RICORRENZA A PARTIRE DA \eqref{clawsalt}: IL PROBLEMA E' CHE NEI CASI INTERESSANTI $d_V^2\ne 0$. IN OGNI CASO E' UN PUNTO SU CUI RIFLETTERE.
\end{remark}
%\end{bf}

\section{Cohomological equation for symmetries and  related recurrence relations}
We have seen that, in canonical coordinates,
 the Christoffels symbols of the compatible connection $\nabla$
 define a semihamiltonian hierarchy. The flow of the hierarchy
\beq
u^i_t=v^i\,u^i_x,\qquad i=1,\dots,n
\eeq
is obtained solving the equation \eqref{sym}:
$$\f{\d_j v^i}{v^j-v^i}=\Gamma^i_{ij}$$
for the unknown characteristic velocities $v^i$.

In this section, we explore the cohomological nature of the equation \eqref{sym}. The results obtained will be important in constructing, in the last Section, a twisted Lenard-Magri chain associated to a semisimple $F$-manifold with eventual identity $E$ and two almost hydrodynamically equivalent connections $\nabla^{(1)}$ and $\nabla^{(2)}$. 

Consider the flows of the hierarchy 
\beq\label{shinv}
u^i_t=V^i_j(u)\,u^j_x=(X\circ)^i_j\,u^j_x,\qquad i=1,\dots,n
\eeq
associated with the semisimple $F$-manifold written in an arbitrary
 coordinate system.

\begin{theorem}\label{equivalences}
1.  The hierarchy \eqref{shinv} is defined by the set of 
 tensor fields $V$ commuting with $L=E\,\circ$ and satisfying the equation
\beq\label{sym1}
(d_{\nabla}V)^i_{jk}=0.
\eeq
Moreover, if $V$ is a tensor field commuting with $L=E\, \circ$, and satisfying $d_{\nabla}V=0$, then we have also $d_{\tilde \nabla} V=0$ for any other hydrodynamically equivalent or almost equivalent connection $\tilde \nabla$ (in this case compatible with $*$). 
 %or almost equivalent connection \textcolor{red}{compatible with $*$}, always assuming the commutativity with $L$\newline
 
2. The hierarchy \eqref{shinv} is defined by the set of  vector fields $X$ satisfying
\beq\label{sym1bis}
d_{\nabla}(X\circ)^i_{jk}=c^i_{jl}\nabla_k X^l-c^i_{kl}\nabla_j X^l=0.
\eeq
In the equation \eqref{sym1bis} one can substitute $\nabla$ with any other hydrodynamically equivalent
 or almost equivalent connection $\tilde\nabla$ and the structure
 constants $c^i_{jk}$ with the structure constants $\tilde c^i_{jk}$
 of the product $*$ compatible with $\tilde\nabla$.
\newline
3. The hierarchy \eqref{shinv} is defined by the set of  vector fields $X$ satisfying
\beq\label{sym3}
[d_{\nabla}X,L]^j_i= (d_{L\nabla}X)^j_{i}-L^j_l(d_{\nabla}X)^l_{i}=0.
\eeq
where $[\cdot,\cdot]$ is the commutator of matrices.
 In the equation \eqref{sym3} one can substitute $\nabla$ with any other hydrodynamically equivalent
 connection. The hierarchy can be also defined by
\beq\label{shinvstar}
u^i_t=V^i_j(u)\,u^j_x=(X\,*)^i_j\,u^j_x,\qquad i=1,\dots,n
\eeq
where the vector fields $X$ are solutions of the equation
\beq\label{sym3star}
[d_{\tilde\nabla}X,L]=0.
\eeq
Here $\tilde\nabla$ is an almost hydrodynamically equivalent connection compatible with $*$.
\end{theorem}

\n
\proof

\n
1. The commutativity with $L=E\, \circ$ tell us that  $V$
 is diagonal in canonical coordinates for $\circ$, $V^i_j=v^j(u)\delta^i_j$, due to the fact that $L$ is diagonal with distinct eigenvalues in these coordinates.  
 Now, since $$(d_{\nabla} V)^i_{jk}=\partial_j V^i_k+\Gamma^i_{lj}V^l_k-\partial_k V^i_j-\Gamma^i_{lk}V^l_j$$
 it is immediate to check, taking into account  \eqref{scc4}, namely the compatibility of $\nabla$ with $\circ$, that for $i\neq j\neq k\neq i$ the above expression vanishes identically. 
 Moreover, from the above expression, setting $k=i\neq j$ one gets in canonical coordinates for $\circ$
 $$0=(d_{\nabla}V)^i_{ji}=\partial_j v^i+\Gamma^i_{ij}v^i-\Gamma^i_{ji}v^j,$$
 which is indeed \eqref{sym} in canonical coordinates. The remaining cases can be treated similarly. 

Now using the commutativity with $L$, we prove that $d_{\nabla} V=0$ implies $d_{\tilde \nabla} V=0$. As we saw above, the commutativity with $L$ and $d_{\nabla} V=0$ in canonical coordinates for $\circ$ reads $\partial_j v^i+\Gamma^i_{ij}v^i-\Gamma^i_{ji}v^j=0$. 

Since $L=E\,\circ=\tilde E\, *$ and $\tilde E^i\neq \tilde E^j$ for $i\neq j$ in canonical coordinates for $*$, due to second point of Theorem \ref{fundamentaltheorem}, we have that $[V,L]=0$ expressed in canonical coordinates for $*$ implies that $V$ is diagonal also in these coordinates, $V=\tilde v^i \delta^i_j$. 

Therefore, since $V$ is diagonal in both the canonical coordinates for $\circ$ and for $*$, we can exploit Remark \ref{importantremark} and by equation \eqref{auxiliaryxxxx} $(d_{\nabla}-d_{\tilde \nabla})V=0$, since $\nabla$ and $\tilde \nabla$ are hydrodynamically almost equivalent. But since $d_{\nabla} V=0$ by assumption,  we get $d_{\tilde \nabla}V=0$.
%If we substitute $\nabla$ with an almost hydrodynamically equivalent connection $\tilde\nabla$
 %nothing changes. Indeed, in canonical coordinates for $*$, one has $\tilde\Gamma^k_{ij}=0$ for $i\ne j\ne k\ne i$
 %and $\tilde\Gamma^i_{ij}=\Gamma^i_{ij}$ for $i\ne j$.
\newline
\newline
2. The tensor fields  of the form
$$V=X\circ$$
clearly commute with $L$ (due to associativity):
$$L^i_k V^k_j=L^i_k c^k_{jl}X^l=c^i_{km}c^k_{jl}X^l E^m=c^i_{kl}c^k_{jm}X^l E^m=L^k_jc^i_{kl}X^l
=L^k_j V^i_k.$$
Therefore for such tensor fields we have only to impose condition \eqref{sym1} that, taking into
 account \eqref{scc}, reduces to
$$
c^i_{jl}\nabla_k X^l=c^i_{kl}\nabla_j X^l,
$$
which is \eqref{sym1bis}. This proves that any solution
 of \eqref{sym1bis} defines a flow of the hierarchy. In fact any flow of the hierarchy can be obtained in this way. Indeed, in canonical coordinates,
 \eqref{sym1bis} is equivalent to the condition
$$\nabla_i X^j=0,\qquad i\ne j.$$
 Using  \eqref{scc3} and \eqref{scc4} we can write the
 above condition as
$$\d_i X^j+\Gamma^j_{ji}(X^j-X^i)=0,\qquad i\ne j$$
and this is exactly condition \eqref{sym} noticing that in canonical coordinates one has the identification $X^i=v^i$.
 Writing the tensor field $V$ as $V=X\,*$ and repeating  the above arguments we can immediately obtain the second claim of point $2$, using the fact that the canonical coordinates for $*$ are of the form $\tilde u^i(u^i)$, due the first claim in Theorem \ref{fundamentaltheorem}.
\newline
\newline
3. In canonical coordinates, taking into accont 
 \eqref{scc3} and \eqref{scc4}, equation \eqref{sym3} reads
\beq\label{sym3bis}
(E^i-E^j)\nabla_i X^j=(E^i-E^j)\left(
\d_i X^j+\Gamma^j_{ji}(X^j-X^i)\right)=0
\eeq
and this is exactly condition \eqref{sym}, since  in canonical coordinate we have the identification $X^i=v^i$ and the components of $E$  are assumed to be functionally independent.

The equation \eqref{sym3} does not change if we substitute $\nabla$ with a hydrodynamically equivalent connection but in general it does change if we substitute $\nabla$ with an almost hydrodynamically equivalent connection $\tilde \nabla$. 
However, we can still prove that tensor fields $X\, *$ define symmetries of the hierarchy. 
Indeed, if $L=E\, \circ$, we can also write $L=\tilde E\, *$ where $\tilde E=E\circ E$ (see the second claim in Theorem \ref{fundamentaltheorem}). 

Thus, since $L=\tilde E \, *$, using the almost hydrodynamically equivalent connection $\tilde \nabla$ and using \eqref{scc3} and \eqref{scc4} in canonical coordinates for $*$ we can rewrite equation \eqref{sym3} as $[d_{\tilde \nabla} X, \tilde E\, *]=0$ or 
\beq\label{sym3tris}
[\tilde E^i-\tilde E^j]\tilde\nabla_i X^j=[\tilde E^i-\tilde E^j]\left(
\d_i X^j+\tilde\Gamma^j_{ji}(X^j-X^i)\right)=0.
\eeq
Since $\tilde E^i=(E\circ E)^i$, we know by the second claim in Theorem \ref{fundamentaltheorem} that in canonical coordinates for $*$ one has $(E\circ E)^i\ne(E\circ E)^j$ and therefore the above equation is equivalent to
$$\d_i X^j+\tilde\Gamma^j_{ji}(X^j-X^i)=0.$$
This means that the tensor fields $X\,*$ define symmetries of the hierarchy.
\endproof

\begin{remark}
From the above theorem it follows that, in the semisimple case, equations \eqref{sym1bis} and  \eqref{sym3} are equivalent. 
In the general case one can prove that \eqref{sym1bis} implies \eqref{sym3}. Indeed
 from \eqref{sym1bis} it follows that
\beq\label{fcons}
e^kc^i_{jl}\nabla_k X^l=e^k c^i_{kl}\nabla_j X^l=\nabla_j X^i
\eeq
and
\beq\label{scons}
E^k c^i_{jl}\nabla_k X^l=E^kc^i_{kl}\nabla_j X^l=L^i_l\nabla_j X^l. 
\eeq
Using \eqref{fcons} we can write the left hand side as
$$E^k c^i_{jl}\nabla_k X^l=e^mE^k c^i_{jl}c^l_{kp}\nabla_m X^p
=e^mE^k c^i_{pl}c^l_{kj}\nabla_m X^p=L^l_je^mc^i_{pl}\nabla_m X^p=
L^l_j\nabla_l X^i.
$$
\end{remark}

\subsection{The flat case: the principal hierarchy}
Let us consider more in the detail the case in which the connection $\nabla$ is flat, namely $d_{\nabla}^2=0$ identically. This automatically implies  the following remarkable fact:
 any solution $X$  of \eqref{sym3} (or of the equivalent  equation \eqref{sym1bis})  defines a solution $V=d_{\nabla}X$ of \eqref{sym1} commuting with $L$
 (and viceversa, since due to the triviality of the cohomology in the flat case,
 any solution of \eqref{sym1} commuting with $L$
 can be obtained in this way).

In the flat case, this means that any solution $X$ of  \eqref{sym3} (or \eqref{sym1bis})
 defines \emph{two} different commuting flows: one given
 by
$$u^i_{t_1}=c^i_{jk}X^k\,u^j_x,\qquad i=1,\dots,n.$$
and one given by
$$u^i_{t_2}=(d_{\nabla}X)^i_{j}\,u^j_x=c^i_{jk}Y^k\,u^j_x,\qquad i=1,\dots,n.$$  
This suggests a recursive procedure to obtain solutions  of \eqref{sym3} (or \eqref{sym1bis}). 

\begin{theorem}
Let $X_{(0)}$ be a solution of \eqref{sym3}/\eqref{sym1bis} then the vector fields defined recursively by
\beq\label{rec1}
(d_{\nabla}X_{(p+1)})^i_k=\nabla_k X_{(p+1)}^i =-c^i_{kl}X^l_{(p)}
\eeq
are still solutions of \eqref{sym3}/\eqref{sym1bis}.
\end{theorem}

\n
{\emph Proof \cite{LPR}}. Suppose $X_{(p)}$ is a solution of \eqref{sym1bis} ($d_{\nabla}(c^i_{jl}X^l_{(p)})=0$). 
This means that the equation \eqref{rec1} admits a solution, let's say $X_{(k+1)}$. It is easy to check that it
 satisfies \eqref{sym1bis}. Indeed
%$$L^k_j\nabla_k X^i_{(p+1)}=L^k_jc^i_{kl}X^l_{(p)}
%=c^k_{jm}c^i_{kl}E^m X^l_{(p)}=c^k_{jl}c^i_{km}E^m X^l_{(p)}=L^i_k\nabla_j X^k_{(p+1)}.$$
$$c^i_{jl}\nabla_k X^l_{(p+1)}=-c^i_{jl}c^l_{km}X^m_{(p)}
=-c^i_{kl}c^l_{jm}X^m_{(p)}=c^i_{kl}\nabla_j X^l_{(p+1)}.$$
The recursion obviously proceeds also in the opposite
 direction (read the proof from the right to the left).
\endproof

 In this way starting from $X_{(0)}$ one can define recursively $X_{(1)},X_{(2)},X_{(3)},\dots$
 and $X_{(-1)},X_{(-2)},X_{(-3)},\dots$. In the negative direction the procedure stops if $\nabla X_{(-k)}=0$
 for some $k$. The above recurrence relations coincide with the recurrence relations of the principal hierarchy.

In the next section we study the case of $F$-manifolds
 with two compatible equivalent flat connections.

\section{Bidifferential calculus and 
 principal hierarchy}\label{recursionprincipal}
With consider now the case of a semisimple $F$-manifold with two compatible hydrodynamically equivalent flat connections $\nabla^{(1)}$ and $\nabla^{(2)}$.

The $(1,1)$ tensor fields $V$ defining the hierarchy  satisfy both the equations
\beq\label{dnav1}
d_{\nabla^{(1)}}V=0
\eeq
and
\beq\label{dnav2}
d_{\nabla^{(2)}}V=0
\eeq
Since the connections are flat the equations \eqref{dnav1} and \eqref{dnav2} imply
$$V=d_{\nabla^{(1)}}X_1=d_{\nabla^{(2)}}X_2,$$
for two suitable  vector fields $X_1$ and $X_2$. 

This means that the vector fields $X_1$ is a solution of 
\beq\label{d2d1}
d_{\nabla^{(2)}}\,d_{\nabla^{(1)}}\,X_1=0.
\eeq
and the vector field $X_2$ is a solution of
\beq\label{d1d2}
d_{\nabla^{(1)}}\,d_{\nabla^{(2)}}\,X_2=0.
\eeq
In general, the differential $d_{\nabla^{(1)}}$ and
 $d_{\nabla^{(2)}}$ do not anticommute and therefore
 we cannot conclude that $X_1$ and $X_2$ are different  solutions of the \emph{same} equation, let's say \eqref{d1d2}. However due to the identity (compare part $3$ of Theorem \ref{equivalences})
\beq\label{meq}
[d_{\nabla^{(1)}}X,L]=[d_{\nabla^{(2)}}X,L].
\eeq
the vanishing of $[d_{\nabla^{(1)}}X,L]$ implies
 the vanishing of $[d_{\nabla^{(2)}}X,L]$ and vice versa. This means that if $V=d_{\nabla^{(1)}}X$
 defines a symmetry then also $V=d_{\nabla^{(2)}}X$
 defines a symmetry and vice versa. In this case
$$d_{\nabla^{(1)}}d_{\nabla^{(2)}}X
=0=d_{\nabla^{(2)}}d_{\nabla^{(1)}}X.$$

Let us consider now a recursive procedure to find
 solutions of \eqref{d1d2} defining symmetries. 
\begin{theorem}\label{27}
Equation \eqref{d1d2} can be solved recursively. More precisely, given a vector field $X_{(p)}$ satisfying \eqref{d1d2}, the vector field
 $X_{(p+1)}$ defined by 
\beq\label{LenardMagri}
d_{\nabla^{(1)}}X_{(p+1)}=d_{\nabla^{(2)}}X_{(p)}.
\eeq
 is a new solution of \eqref{d1d2}. Moreover if
  $d_{\nabla^{(1)}}X_{(p)}$ commutes with $L$ then
also $d_{\nabla^{(1)}}X_{(p+1)}$ commute with $L$.
 \end{theorem}
 \proof
 First of all, let us notice that the above recurrence relations are well defined since the right
 hand side is $d_{\nabla^{(1)}}$-closed. The starting 
 points of the recurrence relations are the flat vector fields of the connection $\nabla^{(1)}$ (obviously
 we can start from  the flat vector fields of the connection $\nabla^{(2)}$ exchanging the role of the
 two connections). We have to check that the tensor
 fields obtained using the above procedure commute
 with $L=E\, \circ$. In other words we have to check that if $[d_{\nabla^{(1)}}X_{(p)},L]=0$ then also
 $[d_{\nabla^{(1)}}X_{(p+1)},L]=0$. But this follows
 from \eqref{LenardMagri} and from the fact, already pointed out, that in the case of hydrodynamically equivalent connections we have the identity
 \eqref{meq}.

\endproof
Summarizing, the proposition above provides us with the following Lenard-Magri chain
\begin{displaymath}
\begin{array}{rcl}
&&0=d_{\nabla^{(1)}}X_{(0,\alpha)}\\
&\stackrel{d_{\nabla^{(1)}}}{\nearrow}&\\
X_{(0,\alpha)}&&\\
& \stackrel{d_{\nabla^{(2)}}}{\searrow} &\\
&& d_{\nabla^{(2)}}X_{(0,\alpha)}=d_{\nabla^{(1)}}X_{(1,\alpha)} \\
&\stackrel{d_{\nabla^{(1)}}}{\nearrow}&\\
X_{(1,\alpha)}&&\\
& \stackrel{d_{\nabla^{(2)}}}{\searrow} &\\
&& d_{\nabla^{(2)}}X_{(1,\alpha)}=d_{\nabla^{(1)}}X_{(2,\alpha)} \\
&\stackrel{d_{\nabla^{(1)}}}{\nearrow}&\\
X_{(2,\alpha)}&&\\
& \stackrel{d_{\nabla^{(2)}}}{\searrow} &\\
&& d_{\nabla^{(2)}}X_{(2,\alpha)}=d_{\nabla^{(1)}}X_{(3,\alpha)} \\
&\stackrel{d_{\nabla^{(1)}}}{\nearrow} & \\
X_{(3,\alpha)}&& \\
&\stackrel{d_{\nabla^{(2)}}}{\searrow} & \\
&& d_{\nabla^{(2)}}X_{(3,\alpha)}=d_{\nabla^{(1)}}X_{(4,\alpha)} \\
&\stackrel{d_{\nabla^{(1)}}}{\nearrow} & \\
X_{(4,\alpha)}&& \\
&\stackrel{d_{\nabla^{(2)}}}{\searrow} & \\
&&  .\,.\,.
\end{array}
\end{displaymath}
where $(X_{(0,1)},\dots,X_{(0,n)})$ is a frame of flat
 vector fields for $\nabla^{(1)}$.  The corresponding  equations of the associated hierarchy are
$$u^i_{t_{(p,\alpha)}}=(d_{\nabla^{(2)}}X_{(p-1,\alpha)})^i_j\,u^j_x=(d_{\nabla^{(1)}}X_{(p,\alpha)})^i_j\,u^j_x,\qquad i=1,\dots,n,\,p=1,2,3,\dots$$
Observe that in this case we have additional symmetries given by the flows
$$u^i_{\tau_{(p,\alpha)}}=(X_{(p,\alpha)}\circ)^i_j\,u^j_x,\qquad i=1,\dots,n,\,p=0,1,2,\dots$$

\subsection{An important example}
In the case of an $F$-manifold with compatible flat connection $\nabla^{(1)}$ we can choose as second flat connection $\nabla^{(2)}$ one of the connections of the one-parameter family:
$$\tilde\Gamma^i_{jk}=\Gamma^i_{jk}+zc^i_{jk}.$$
Let us observe that each new connection is hydrodynamically equivalent
 to the old one. Indeed in canonical coordinates 
 the difference between the two connections, namely
 the term $zc^i_{jk}$, is different from zero if and only if all the indices are equal.
It is interesting to compare the Lenard-Magri chains corresponding
 to the choice $z=-1$ with the recursion relations of
 the principal hierarchy \eqref{rec1}. The former can be written as
\beq
\nabla_j (X_{(p+1)}-X_{(p)})^i =-c^i_{jl}X^l_{(p)},
\eeq
while the latter are
\beq
\nabla_j Z_{(p+1)}^i =-c^i_{jl}Z^l_{(p)}.
\eeq
Now it turns out that starting from the same vector field, the vector fields $X_{(p)}$ obtained through the Lenard-Magri chain are just a linear combination of those obtained via the recursion relations of the principal hierarchy $Z_{(l)}$ for $0\leq l\leq p$. This is the meaning of the following:
\begin{proposition}
Let $X_{(k)}$  be the vector fields obtained using the recursion relations for the Lenard-Magri chains corresponding to the choice $z=-1$ and let $Z_{(k)}$ be the vector fields constructed using the recursion relations for the principal hierarchy. Then if the two systems of recursion relations start at the same point $Z_{(0)}=X_{(0)},$ we have that 
 the vector fields $X_{(p)}$ can be written explicitly
 in terms of the vector fields $Z_{(l)}$ as
$$X_{(p)}=\sum_{k=0}^p\binom{p}{k}Z_{(k)}.$$
\end{proposition}
\proof
The proof is a straightforward computation:
\begin{eqnarray*}
\nabla_j(X_{(n+1)}-X_{(n)})^i&=&\nabla_j\left\{\sum_{k=0}^{n+1}
\binom{n+1}{k}Z^i_{(k)}-\sum_{k=0}^n\binom{n}{k}Z^i_{(k)}\right\}=\\
&=&\nabla_j
\left\{Z^i_{(n+1)}+\sum_{k=1}^{n}
\left[\binom{n+1}{k}-\binom{n}{k}\right]Z^i_{(k)}\right\}=\\
&=&\nabla_j
\left\{Z^i_{(n+1)}+\sum_{k=1}^{n}
\binom{n}{k-1}Z^i_{(k)}\right\}=\\
&=&-c^i_{jl}\left\{Z^l_{(n)}+\sum_{k=1}^{n}
\binom{n}{k-1}Z^l_{(k-1)}\right\}=\\
&=&-c^i_{jl}\left\{Z^l_{(n)}+\sum_{k=0}^{n-1}
\binom{n}{k}Z_{(k)}^l\right\}=\\
&=&-c^i_{jl}\left\{\sum_{k=0}^{n}
\binom{n}{k}Z^l_{(k)}\right\}=-c^i_{jl}X^l_{(n)}
\end{eqnarray*}
\endproof

\begin{remark}
Due to \eqref{F2} and \eqref{scc}, the pencil of connections  
$\nabla^{(\lambda)}=\nabla^{(2)}-\lambda\nabla^{(1)}$
in the above example is flat for every $\lambda$. As a consequence
 the differentials $d_{\nabla^{(1)}}$ and $d_{\nabla^{(2)}}$ anticommute and we have a
 \emph{bidifferential complex} or, in the language of \cite{DM}, a \emph{bidifferential calculus}.
\end{remark}

\section{Twisted Lenard-Magri chains and bi-Hamiltonian recursion relations}
Let us consider a semisimple $F$-manifold endowed with two almost hydrodynamically equivalent flat connections $\nabla^{(1)}$ and $\nabla^{(2)}$. In the case of a Frobenius manifold, $\nabla^{(1)}$ is the Levi-Civita connection associated with the invariant metric $\eta$  and  $\nabla^{(2)}$ is the Levi-Civita connection associated with the intersection form $g$. 

In this section, we adapt the construction of the previous section to this new situation. This will provide us with a {\em twisted} system of Lenard-Magri chains that under specific conditions reduce to the classical system of Lenard-Magri chains but which, in general, is different. 
\newline
\newline
Like in the case of hydrodynamically equivalent connections, 
the $(1,1)$-tensor fields $V$ defining the hierarchy  satisfy both the equations \eqref{dnav1} and \eqref{dnav2} and therefore
$$V=d_{\nabla^{(1)}}X_1=d_{\nabla^{(2)}}X_2,$$
for two suitable  vector fields $X_1$ and $X_2$. It is also still true that if $X_{(p)}$ is a solution of \eqref{d1d2} then the Lenard-Magri relation 
$$
d_{\nabla^{(1)}}X_{(p+1)}=d_{\nabla^{(2)}}X_{(p)}.
$$
defines correctly $X_{(p+1)}$. However the next 
 step of the recursion 
$$
d_{\nabla^{(1)}}X_{(p+2)}=d_{\nabla^{(2)}}X_{(p+1)}.
$$
is not well defined since, by construction, $X_{(p+1)}$
 is a solution of \eqref{d2d1} \emph{but not} of
 \eqref{d1d2}, in general. The problem is that the identity \eqref{meq} is no longer satisfied for almost hydrodynamically equivalent connections.
 For the same reason if $[d_{\nabla^{(2)}}X_{(p)},L]=0$ then  $[d_{\nabla^{(1)}}X_{(p+1)},L]=0$, but, in general
 $[d_{\nabla^{(2)}}X_{(p+1)},L]\ne 0$. In order to
 keep the recursion we need an additional step
 mapping solutions of $[d_{\nabla^{(1)}}X,L]=0$
 (and therefore of \eqref{d2d1}) 
 into solutions of $[d_{\nabla^{(2)}}X,L]=0$
 (and therefore of \eqref{d1d2}) .
 As we will see in a moment this step consists in substituting $X_{(p)}$
 with $E\circ X_{(p)}$, giving rise to a twisted system of Lenard-Magri chains.

\begin{theorem}
Let $M$ be a semisimple $F$-manifold with eventual identity $E$ and two almost hydrodynamically equivalent flat connections $\nabla^{(1)}$ and $\nabla^{(2)}$ (the first one compatible with $ \circ$ and the
 second one compatible with $*$) . Then the following twisted version of the Lenard-Magri chain holds: 
\begin{displaymath}
\begin{array}{rcl}
&&0=d_{\nabla^{(1)}}X_{(0,\alpha)}\\
&\stackrel{d_{\nabla^{(1)}}}{\nearrow}&\\
X_{(0,\alpha)}&&\\
\downarrow &\\
E\circ X_{(0,\alpha)}&&\\
& \stackrel{d_{\nabla^{(2)}}}{\searrow} &\\
&& d_{\nabla^{(2)}}\left(E\circ X_{(0,\alpha)}\right)=d_{\nabla^{(1)}}X_{(1,\alpha)} \\
&\stackrel{d_{\nabla^{(1)}}}{\nearrow}&\\
X_{(1,\alpha)}&&\\
\downarrow &\\
E\circ X_{(1,\alpha)}&&\\
& \stackrel{d_{\nabla^{(2)}}}{\searrow} &\\
&& d_{\nabla^{(2)}}\left(E\circ X_{(1,\alpha)}\right)=d_{\nabla^{(1)}}X_{(2,\alpha)} \\
&\stackrel{d_{\nabla^{(1)}}}{\nearrow}&\\
X_{(2,\alpha)}&&\\
\downarrow &\\
E\circ X_{(2,\alpha)}&&\\
& \stackrel{d_{\nabla^{(2)}}}{\searrow} &\\
&& d_{\nabla^{(2)}}\left(E\circ X_{(2,\alpha)}\right)=d_{\nabla^{(1)}}X_{(3,\alpha)} \\
&\stackrel{d_{\nabla^{(1)}}}{\nearrow} & \\
X_{(3,\alpha)}&& \\
\downarrow &\\
E\circ X_{(3,\alpha)}&&\\
&\stackrel{d_{\nabla^{(2)}}}{\searrow} & \\
&&  .\,.\,.
\end{array}
\end{displaymath}
where $(X_{(0,1)},\dots,X_{(0,n)})$ is a frame of flat
 vector fields. The corresponding equations of the associated hierarchy are
\beq\label{thy}
u^i_{t_{(p,\alpha)}}= [d_{\nabla^{(2)}}\left(E\circ X_{(p-1,\alpha)}\right)]^i_j\,u^j_x=      (d_{\nabla^{(1)}}X_{(p,\alpha)})^i_j\,u^j_x,\qquad i=1,\dots,n,\,p=1,2,3,\dots.
\eeq
\end{theorem}
\proof
If $L=E\,\circ$,
we have to check that if the vector field $X$ satisfies
\beq\label{meqI}
[d_{\nabla^{(1)}}X,L]=0
\eeq 
then the vector field $Z=E\circ X$ satisfies the equation 
\beq\label{meqII}
[d_{\nabla^{(2)}}Z,L]=0.
\eeq
The crucial point is the identity
\begin{equation}\label{auxiliaryiv}
Z\,*=X\circ,
\eeq
which is immediate to check.
Since $V=X\circ$ defines a symmetry we have
$$d_{\nabla^{(1)}}(Z\,*)=d_{\nabla^{(2)}}(Z\,*)=0.$$ 
In particular, taking into account \eqref{scc}, $d_{\nabla^{(2)}}(Z\,*)=0$ gives
$\tilde c^{i}_{jl}\nabla_k^{(2)}Z^l-\tilde c^{i}_{kl}\nabla_j^{(2)}Z^l=0$, where $\tilde c$ are the structure constants of $*$. Now in the canonical coordinates for $*$ this expression is equivalent to $\nabla^{(2)}_i Z^j=0$, $i\neq j$. Indeed it reduces to $\delta^i_j \nabla^{(2)}_k Z^i-\delta^i_k\nabla^{(2)}_j Z^i=0.$ This equation is automatically satisfied for $i=j=k$, and for $i\neq j\neq k\neq i$, while for $j=i\neq k$ it gives $\nabla^{(2)}_k Z^i=0$. The other cases can be treated similarly. 

Moreover, using canonical coordinates for $*$, one finds that $L^i_j=(E\circ E)^i\delta^i_j$ (see the second claim of Theorem \ref{fundamentaltheorem}) and in this system of coordinates equation \eqref{meqII} reads 
$$\nabla^{(2)}_jZ^i(E\circ E)^j\delta^j_k-(E\circ E)^i\delta^i_j \nabla^{(2)}_k Z^j=0.$$
or equivalently 
$$[(E\circ E)^k-(E\circ E)^i]\nabla^{(2)}_k Z^i=0.$$
\newline
To conclude the proof we have to show that the recurrence relations 
\beq\label{NSLM}
d_{\nabla^{(1)}}X_{(p+1,\alpha)}=d_{\nabla^{(2)}}\left(E\circ X_{(p,\alpha)}\right)
\eeq
are well defined, that is
$$d_{\nabla^{(1)}}d_{\nabla^{(2)}}\left(E\circ X_{(p,\alpha)}\right)=0.$$

In order to show this, it is enough to show that the tensor field $V=d_{\nabla^{(2)}}\left(E\circ X_{(p,\alpha)}\right)$ defines a symmetry. 
But by construction, $V$ commutes with $L$ and satisfied $d_{\nabla^{(2)}}V=0$, since $\nabla^{(2)}$ is flat. Therefore by the first point of Theorem \ref{equivalences} we have that $d_{\nabla^{(1)}}V=0$ since $\nabla^{(1)}$ and $\nabla^{(2)}$ are hydrodynamically almost equivalent, and this gives the compatibility of the recurrence relations.  

\endproof

\begin{remark}
Notice that the solutions $X$ of the equation \eqref{meqII} provide symmetries of the semihamiltonian hierarchy in two ways similarly to what happens with the solutions of the equation \eqref{meqI}.
 First of all as
$$u^i_{\tau}=(d_{\nabla^{(2)}}X)^i_j\,u^j_x,\qquad i=1,\dots,n.$$
Indeed the $(1,1)$-tensor field $V=d_{\nabla^{(2)}}X$ is diagonal in canonical coordinates for $*$; indeed this follows from 
  the commutativity with $L=E\,\circ=(E\circ E)\;*$, because by the second claim of Theorem \ref{fundamentaltheorem}, $L$ is still diagonal with distinct eigenvalues in canonical coordinates for $*$, and because condition $d_{\nabla^{(2)}}V=0$, which is satisfied by construction, reduces to \eqref{sym} in canonical coordinates for $*$.

Secondly as
$$u^i_{\tau}=(X\,*)^i_j\,u^j_x,\qquad i=1,\dots,n.$$
Therefore besides the symmetries \eqref{thy} we have also the symmetries
$$u^i_{\tau_{(p,\alpha)}}=(X_{(p,\alpha)}\,\circ)^i_ju^j_x=(X_{(p+1,\alpha)}\,*)^i_j\,u^j_x,\qquad i=1,\dots,n,\,p=0,1,2,\dots$$
\end{remark}

To conclude this section,  let us consider what happens when the almost hydrodynamically equivalent connections are associated with a flat
 pencil of metrics.
 \begin{proposition}
 Let $M$ be an $F$-manifold endowed with eventual identity $E$ and with almost hydrodynamically equivalent connections $\nabla^{(1)}$ and $\nabla^{(2)}$. If $\nabla^{(1)}$ and $\nabla^{(2)}$ are associated with a flat pencil of metrics, then the twisted Lenard-Magri chain \eqref{NSLM} coincides with the classical bihamiltonian Lenard-Magri chain.
 \end{proposition}
 \proof
 In this case, using the flat pencil of metrics, the classical bi-Hamiltonian recursion relations of Lenard-Magri type \eqref{rrcl3} can be written in the form
\beq
g_{(1)}^{jl}\nabla^{(1)}_k dh_l^{(p+1)}=
g_{(2)}^{jl}\nabla^{(2)}_k dh_l^{(p)}
\eeq
or
\beq\label{CLM}
\nabla^{(1)}_k\left(g_{(1)}^{jl} dh_l^{(p+1)}\right)=
\nabla^{(2)}_k\left(g_{(2)}^{jl} dh_l^{(p)}\right),
\eeq
since $\nabla^{(i)}$ is the Levi-Civita connection associated to $g_{(i)}$, $i=1,2$. 

If we identify 
 the vector field $X_{(p)}$ with the vector field $g_{(1)}^{jl} dh_l^{(p)}$ we can write \eqref{CLM}
 as
\beq
\nabla^{(1)}_k\left(X^j_{(p+1)}\right)=
\nabla^{(2)}_k\left(L^j_k X^k_{(p)}\right).
\eeq
This is just the recursion relation \eqref{NSLM}. 
\endproof

\end{document}